\documentclass[twocolumn,longauth]{aa}

\usepackage{aas_macros}
\usepackage{natbib}
\usepackage[varg]{txfonts}
\usepackage{balance}
\usepackage{pgfplots}
\usepgfplotslibrary{external} 
\usepackage{amsmath, amssymb}
\usepackage{siunitx}
\usepackage{graphicx}   
\usepackage{layouts}
\usepackage{color}     
\usepackage{subcaption}  
\usepackage{hyperref}   
\usepackage[capitalise]{cleveref}
\usepackage{import}

\bibpunct{(}{)}{;}{a}{}{,} % to follow the A&A style

\defcitealias{2020A&A...633A.150H}{HE20}
\newcommand{\run}{adapted_2}
\newcommand{\imagepath}{plots/\run /}
\newcommand{\imagepathstatic}{plots/static/}

\begin{document}

\title{The Galactic Faraday rotation sky 2020}
\author{
Sebastian Hutschenreuter \inst{\ref{inst:radboud},\ref{inst:mpa},\ref{inst:lmu}} \and 
Craig S. Anderson \inst{\ref{inst:jansky_at_nrao}}   \and
Sarah Betti \inst{\ref{inst:amherst}} \and
Geoffrey C. Bower \inst{\ref{inst:hilo}} \and 
Jo-Anne Brown  \inst{\ref{inst:calgary}}\and
Marcus Br\"uggen \inst{\ref{inst:hamburg}} \and
Ettore Carretti \inst{\ref{inst:INAF-bologna}} \and
Tracy Clarke \inst{\ref{inst:navy}} \and
Andrew Clegg \inst{\ref{inst:google}} \and
Allison Costa \inst{\ref{inst:virginia}} \and
Steve Croft \inst{\ref{inst:berkeley}, \ref{inst:seti}} \and
Cameron Van Eck \inst{\ref{inst:toronto}} \and
B. M. Gaensler  \inst{\ref{inst:toronto}} \and
Francesco de Gasperin \inst{\ref{inst:hamburg}, \ref{inst:INAF-bologna}} \and
Marijke Haverkorn \inst{\ref{inst:radboud}} \and 
George Heald \inst{\ref{inst:csiro}}\and
Charles L.H. Hull \inst{\ref{inst:naoj}, \ref{inst:alma}, \ref{inst:naoj_fellow}} \and 
Makoto Inoue \inst{\ref{inst:taipei}} \and
Melanie Johnston-Hollitt \inst{\ref{inst:curtin}} \and
Jane Kaczmarek \inst{\ref{inst:dominion}} \and
Casey Law \inst{\ref{inst:caltech}} \and
Yik Ki Ma \inst{\ref{inst:bonn}, \ref{inst:canberra}} \and
David MacMahon \inst{\ref{inst:berkeley}} \and
Sui Ann Mao \inst{\ref{inst:bonn}} \and
Christopher Riseley \inst{\ref{inst:INAF-bologna}, \ref{inst:bologna}} \and
Subhashis Roy \inst{\ref{inst:pune}}  \and
Russell Shanahan \inst{\ref{inst:calgary}} \and
Timothy Shimwell \inst{\ref{inst:astron}, \ref{inst:leiden}} \and
Jeroen Stil \inst{\ref{inst:calgary}} \and
Charlotte Sobey \inst{\ref{inst:csiro}} \and
Shane P. O'Sullivan \inst{\ref{inst:dublin}} \and
Cyril Tasse \inst{\ref{inst:obs_paris}, \ref{inst:grahamstown}} \and
Valentina Vacca \inst{\ref{inst:INAF-cagliari}} \and
Tessa Vernstrom \inst{\ref{inst:csiro}} \and
Peter K.G. Williams \inst{\ref{inst:harvard}} \and
Melvyn Wright \inst{\ref{inst:berkeley}} \and
Torsten A. En{\ss}lin \inst{\ref{inst:mpa},\ref{inst:lmu}}
}
\institute{
Department of Astrophysics/IMAPP, Radboud University, P.O. Box 9010,6500 GL Nijmegen, The Netherlands \label{inst:radboud} 
\and 
Max Planck Institute for Astrophysics, Karl-Schwarzschildstr.1, 85741 Garching, Germany \label{inst:mpa} 
\and 
Ludwig-Maximilians-Universit\"at M\"unchen, Geschwister-Scholl-Platz 1, 80539 Munich, Germany \label{inst:lmu} 
\and 
Jansky fellow of the National Radio Astronomy Observatory, 1003 Lopezville Rd, Socorro, NM 87801 USA 
\label{inst:jansky_at_nrao} 
\and
Department of Astronomy, University of Massachusetts, Amherst, 710 North Pleasant Street, Amherst, MA 01003, USA \label{inst:amherst} 
\and
Academia Sinica Institute of Astronomy and Astrophysics, 645 N. A'ohoku Place, Hilo, HI 96720, USA \label{inst:hilo}
\and
Department of Physics and Astronomy, University of Calgary, 2500 University Drive NW, Calgary AB T2N 1N4, Canada \label{inst:calgary}
\and
Hamburger Sternwarte, Universi\"at Hamburg, Gojenbergsweg 112, 21029 Hamburg, Germany 
\label{inst:hamburg} 
\and
INAF - Istituto di Radioastronomia, via P. Gobetti 101, 40129, Bologna, Italy \label{inst:INAF-bologna} 
\and
U.S. Naval Research Lab, 4555 Overlook Ave., SW Washington, DC 20375, USA \label{inst:navy} 
\and
Google LLC, 1900 Reston Metro Plaza, Reston, VA 20190, USA\label{inst:google}
\and
Department of Astronomy, University of Virginia, Charlottesville, VA 22904, USA  \label{inst:virginia}
\and
Radio Astronomy Lab, 501 Campbell Hall {\#}3411, University of California, Berkeley, CA 94720, USA \label{inst:berkeley}
\and
SETI Institute, 189 N Bernardo Ave {\#}200, Mountain View, CA 94043, USA \label{inst:seti}
\and
Dunlap Institute for Astronomy and Astrophysics, University of Toronto, 50 St. George Street, Toronto, ON M5S 3H4, Canada \label{inst:toronto}
\and
CSIRO Astronomy and Space Science, PO Box 1130, Bentley WA 6102, Australia \label{inst:csiro} 
\and
National Astronomical Observatory of Japan, NAOJ Chile, Alonso de C{\'o}rdova 3788, Office 61B, 7630422, Vitacura, Santiago, Chile \label{inst:naoj} 
\and
Joint ALMA Observatory, Alonso de C{\'o}rdova 3107, Vitacura, Santiago, Chile \label{inst:alma}
\and
NAOJ Fellow \label{inst:naoj_fellow}
\and
Academia Sinica Institute of Astronomy and Astrophysics, ASMAB, NTU, No.1, Roosevelt Rd., Sec. 4, Taipei 10617, Taiwan, R.O.C. \label{inst:taipei}
\and
International Centre for Radio Astronomy Research (ICRAR), Curtin University, Bentley, WA 6102, Australia \label{inst:curtin}
\and
Dominion Radio Astrophysical Observatory, National Research Council of Canada, Box 248, Penticton, BC, V2A 6J9, Canada \label{inst:dominion}
\and
Department of Astronomy and Owens Valley Radio Observatory, California Institute of Technology, Pasadena CA 91125, USA \label{inst:caltech}
\and
Max-Planck-Institut f\"ur Radioastronomie, Auf dem H\"ugel 69, 53121 Bonn, Germany \label{inst:bonn}
\and
Research School of Astronomy \& Astrophysics, Australian National University, Canberra, ACT 2611, Australia \label{inst:canberra} 
\and
Dipartimento di Fisica e Astronomia, Universit{\`a} degli Studi di Bologna, via P. Gobetti 93/2, 40129 Bologna, Italy \label{inst:bologna}
\and
National Centre for Radio Astrophysics, TIFR, Pune University Campus, Pune, India \label{inst:pune}
\and
ASTRON, Netherlands Institute for Radio Astronomy, Oude Hoogeveensedijk 4, 7991 PD, Dwingeloo, The Netherlands \label{inst:astron}
\and
Leiden Observatory, Leiden University, PO Box 9513, NL-2300 RA Leiden, The Netherlands \label{inst:leiden}
\and
School of Physical Sciences and Centre for Astrophysics \& Relativity, Dublin City University, Glasnevin D09 W6Y4, Ireland \label{inst:dublin}
\and
GEPI{\&}USN, Observatoire de Paris, CNRS, Universit\'e Paris Diderot, 5 place Jules Janssen, 92190 Meudon, France \label{inst:obs_paris}
\and
Centre for Radio Astronomy Techniques and Technologies, Department of Physics and Electronics, Rhodes University, Grahamstown 6140, South Africa \label{inst:grahamstown} 
\and
INAF - Osservatorio Astronomico di Cagliari, Via della Scienza 5, 09047 Selargius (CA), Italy \label{inst:INAF-cagliari} 
\and
Harvard-Smithsonian Center for Astrophysics, 60 Garden St. MS-20, Cambridge, MA 01238, USA  \label{inst:harvard}
}

\abstract {}{This work gives an update to existing reconstructions of the Galactic Faraday rotation sky by processing almost all Faraday rotation data sets available at the end of the year 2020. 
Observations of extra-Galactic sources in recent years have, among other regions, further illuminated the previously under-constrained southern celestial sky, as well as parts of the inner disc of the Milky Way.
This has culminated in an all-sky data set of 55,190 data points, which is a significant expansion on the 41,330 used in previous works, hence making an updated separation of the Galactic component a promising venture.   
The increased source density allows us to present our results in a resolution of about $1.3\cdot 10^{-2}\, \mathrm{deg}^2$ ($46.8\,\mathrm{arcmin}^2$), which is a twofold increase compared to previous works.
}{
As for previous Faraday rotation sky reconstructions, this work is based on information field theory, a Bayesian inference scheme for field-like quantities which handles noisy and incomplete data. 
}{
In contrast to previous reconstructions, we find a significantly thinner and pronounced Galactic disc with small-scale structures exceeding values of several thousand $\mathrm{rad}\,\mathrm{m}^{-2}$. 
The improvements can mainly be attributed to the new catalog of Faraday data, but are also supported by advances in correlation structure modeling within numerical information field theory.
We furthermore give a detailed discussion on statistical properties of the Faraday rotation sky and investigate correlations to other data sets.
}{}

\maketitle

\section{Introduction}
\label{sec:intro}
The Faraday effect describes the rotation of the polarization position angle propagating through a magnetized plasma, and provides information on the line-of-sight (LOS) component of magnetic fields weighted by the thermal electron density. 
These quantities are key puzzle pieces for the characterization and modeling of the structure of the Milky Way's magnetic field and of many extra-Galactic astrophysical objects, such as galaxies \citep{2015A&ARv..24....4B} and galaxy clusters \citep{2013A&ARv..21...62D}. A better understanding of these structures is not only an interesting topic by its own merit but may also provide a pathway in understanding the origin of the magnetic field in the Universe as a whole \citep{2016RPPh...79g6901S}. 
A primary objective of research on cosmic magnetism has therefore been the investigation of the polarimetric properties of peculiar extra-Galactic objects, in the hope to obtain insights on their morphology and formation history.\par
Over the years, this pursuit has led to a dense coverage of Faraday rotation data across the whole sky, which have been cataloged in exhaustive compilations \citep{2012A&A...542A..93O, van_Eck_catalog}.  
The characteristics of this data not only depend on the properties of the objects in question but on the integral of all environments through which the light has passed upon arriving at the telescope. 
This leads to a potential intertwining of information of completely different astrophysical environments, such as the intergalactic and interstellar media.  
In  particular, the plasma of our Galaxy is responsible for a significant amount of Faraday rotation, see \citet{1962Natur.195.1084C} or \citet{1966AuJPh..19..129G} for early detections of the Galactic Faraday rotation contribution and \citet{2015ASSL..407..483H} and \citet{2015A&ARv..24....4B} for reviews.\par
In order to disentangle the contributions to Faraday rotation from multiple sources, reconstruction procedures have been developed. 
Specifically in the case of the Galactic Faraday rotation sky, a useful measure for discerning Galactic and extra-Galactic components is the similarity of RM values of nearby sources on the sky. 
One expects that large angular correlations in Faraday data more likely result from local effects, implying that data resulting from extra-Galactic processes should be mostly uncorrelated on angular scales of arc-minutes or larger \citep{2010ApJ...723..476A}.
Exceptions to this might be caused by neighboring parts of the cosmic large-scale structure or large-scale magnetic fields in the inter-galactic medium in proximity to the Milky Way \cite{ 2006ApJ...637...19X}. 
Past all-sky methods to reconstruct the Galactic Faraday sky were developed for instance by \citet{2001MNRAS.325..649F, 2004mim..proc...13J, 2005MNRAS.362..403D, 2006ApJ...637...19X,2007BayAn...2..665S, 2011ApJ...738..192P, 2012A&A...542A..93O, 2014RAA....14..942X}.\par
In this paper, we are continuing the work of \citet{2012A&A...542A..93O}, \citet{2015A&A...575A.118O} and \citet{2020A&A...633A.150H}. In the latter reference, henceforth abbreviated with \citetalias{2020A&A...633A.150H}, the authors introduce two inference models. 
The simpler one is a generalization of the model used by \citet{2012A&A...542A..93O}, while the more complicated one additionally restricts the amplitude of the Faraday sky with data on the emission measure (EM) of thermal electrons, as obtained from the \textit{Planck} satellite \citep{2016A&A...594A..10P}.
These works use Bayesian inference schemes, which utilize the whole-sky correlation structure of the Galactic part of the Faraday rotation sky, in order to (a) perform the aforementioned component separation and (b) to interpolate between data points on the sky. 
As the precise form of the correlation structure (which is necessary for the interpolation) is unknown, it is also inferred jointly with the sky map. \par
The aim of this paper is to provide the community with the most recent version of the Galactic Faraday sky with a minimal set of physical assumptions in unprecedented resolution and to draw attention to previously under-constrained sky regions, such as the Magellanic Clouds or peculiar substructures of the Galactic disc. 
The more complicated inferences using the algorithms developed in \citetalias{2020A&A...633A.150H} have also been repeated. 
However, discussing these results, in particular in light of the many options that exist for the inclusion of new data sets and modeling assumptions, exceeds the scope of this paper, which is to provide a methodologically clean update on the Faraday rotation sky with a specific focus on high resolution, specific Galactic structures and statistical properties of the Faraday rotation sky.  
Thus, we perform the same analysis as in \citetalias{2020A&A...633A.150H}, but are restricting ourselves to the model which uses Faraday data only. 
An accompanying publication is in preparation \citep{HE_in_prep}, which focuses on the aforementioned more complicated models and focuses on the analysis of the component maps.
This present publication should therefore be regarded as the primary reference for the updated Faraday rotation sky, whereas the other as providing a physical interpretation of the component maps, which depends more on assumptions. 
The resulting posterior mean and uncertainty maps for the Faraday rotation sky are available for download either as full-sky maps\footnote{\href{https://wwwmpa.mpa-garching.mpg.de/~ensslin/research/data/faraday2020.html}{https://wwwmpa.mpa-garching.mpg.de/{\textasciitilde}ensslin/research/data/faraday2020.html}} or via a cutout server\footnote{\href{http://cutouts.cirada.ca/rmcutout}{CIRADA cutout server}}.\par
We structure the paper as following:
\cref{sec:model} summarizes the relevant physics and the modeling. \cref{sec:data} describes the data used in this work.
\cref{sec:results} discusses the results and \cref{sec:conclusion} gives a conclusive summary. 
 
\begin{figure}[h!]
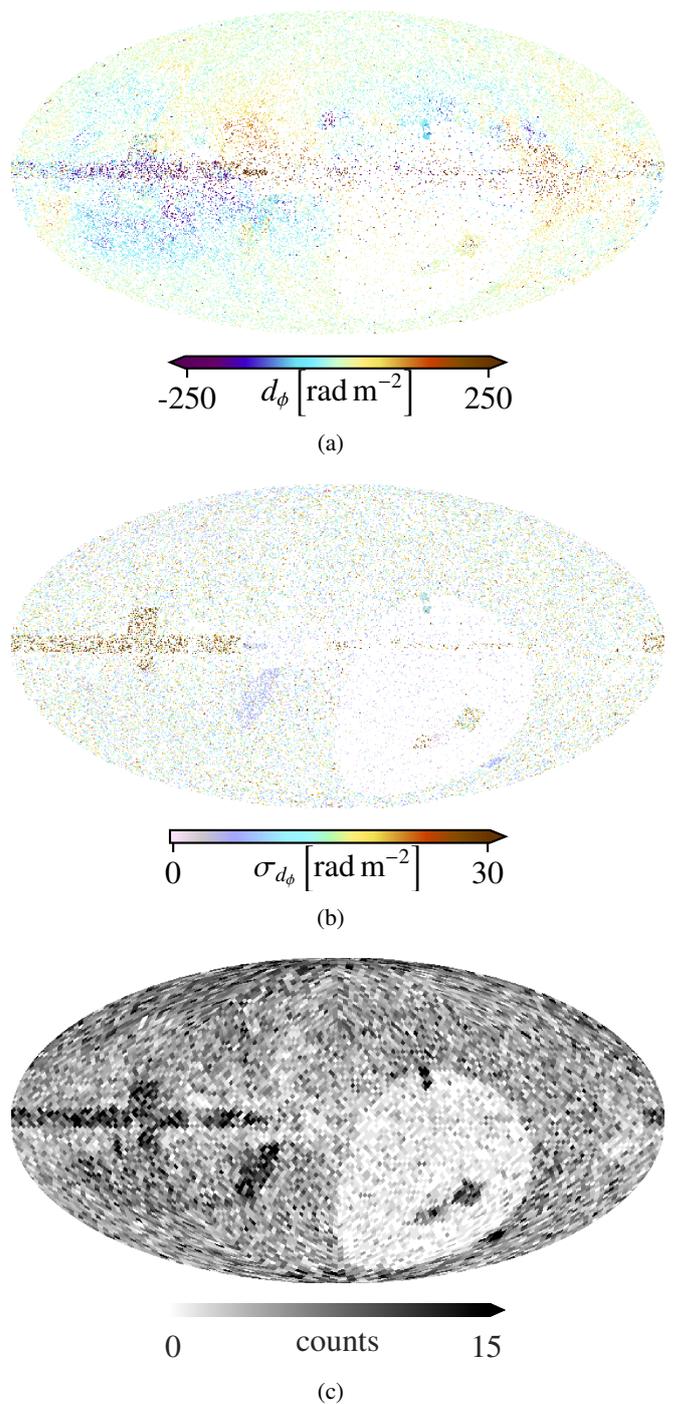

\centering
\begin{subfigure}{0.49\textwidth}
\import{\imagepathstatic}{faraday_8d_faraday_data.pgf}
\caption{\label{fig:faraday_data_mean}}
\end{subfigure}
\begin{subfigure}{0.49\textwidth}
\import{\imagepathstatic}{faraday_8d_faraday_stddev.pgf}
\caption{\label{fig:faraday_data_std} }
\end{subfigure}
\begin{subfigure}{0.49\textwidth}
\import{\imagepathstatic}{faraday_8d_faraday_stddev_counts.pgf}
\caption{\label{fig:faraday_counts} }
\end{subfigure}
\caption{\label{fig:faraday_data} Sky projections of the data set used in this work, the associated observational errors and the source density. 
\Cref{fig:faraday_data_mean} shows the Faraday data, \cref{fig:faraday_data_std} the observed error bars. 
Data points falling into the same pixel have been averaged. 
These plot has been made with a four-times decreased resolution (corresponding to an pixel area of $0.21\,\mathrm{deg}^2$ or $7.56\cdot 10^{2}\,\mathrm{arcmin}^2$ ) compared to that of the inference in order to enable better visualization. 
\Cref{fig:faraday_counts} shows the source density of the data set used in this work. 
The pixels in this map have an area of about $3.36\,\mathrm{deg}^2$ or $1.2 \cdot 10^{4}\,\mathrm{arcmin}^2$, again for visualization purposes.
These, and all subsequent maps, are presented in Galactic coordinates centered at $(l,b) = \left(\ang{0}, \ang{0}\right)$.
}
\end{figure}

\begin{figure*}
\begin{subfigure}{\textwidth}
\centering
\import{\imagepath}{\run _faraday_sky_mean_250.pgf}
\caption{\label{fig:results_250_mean}}
\end{subfigure}
\begin{subfigure}{\textwidth}
\import{\imagepath}{\run _faraday_sky_std_80.pgf}
\caption{\label{fig:results_80_std}}
\end{subfigure}
\caption{\label{fig:results_25080}
Inference results for the posterior mean (\cref{fig:results_250_mean}) and uncertainties  (\cref{fig:results_80_std}) of the Galactic Faraday rotation sky using the updated data catalog. The color scale is saturated at $\pm 250\, \mathrm{rad}\,\mathrm{m}^{-2}$ for the mean and at $80\, \mathrm{rad}\,\mathrm{m}^{-2}$ for the uncertainties to highlight mid latitude features.
}
\end{figure*}
\begin{figure*}
\begin{subfigure}{\textwidth}
\centering
\import{\imagepath}{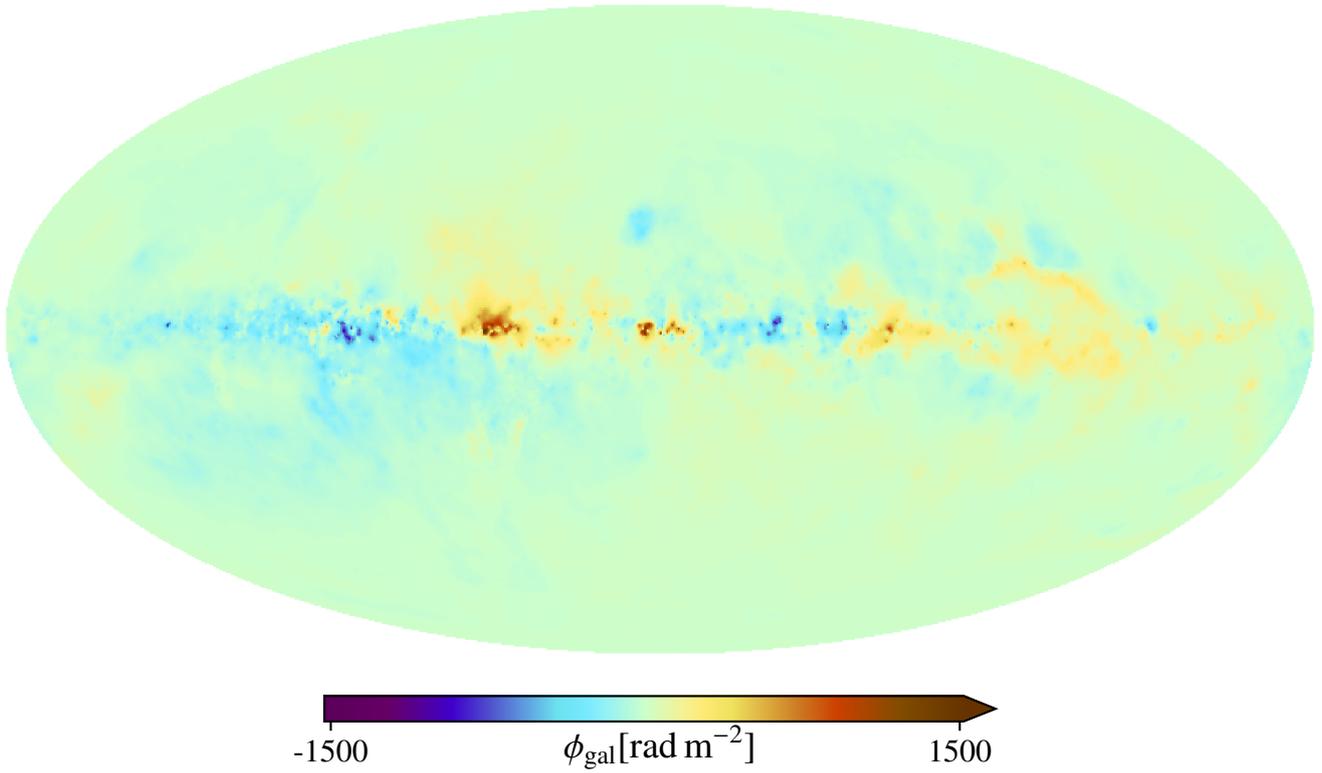}
\caption{\label{fig:results_1500_mean}}
\end{subfigure}
\begin{subfigure}{\textwidth}
\import{\imagepath}{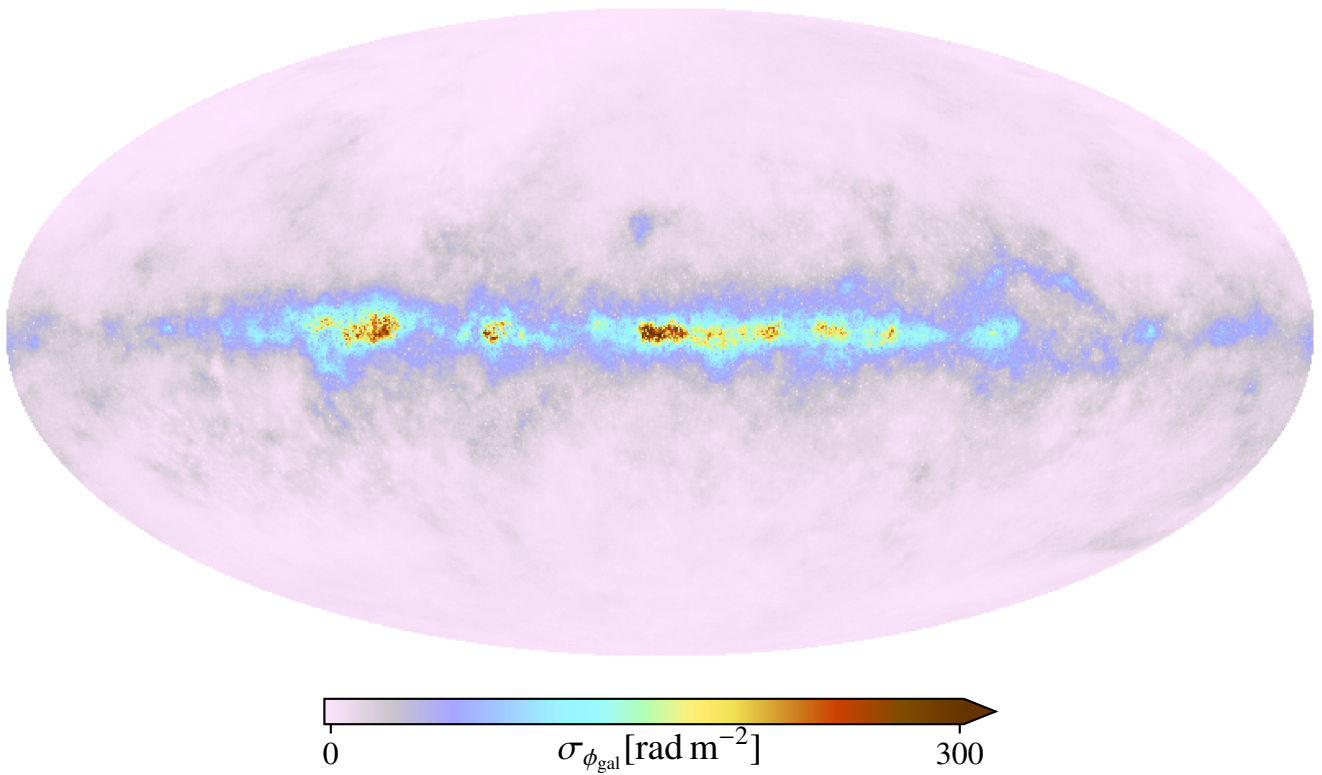}
\caption{\label{fig:results_300_std}}
\end{subfigure}
\caption{\label{fig:results_1500300}  
As for \cref{fig:results_25080}, but the color scale is saturated at $\pm 1500\, \mathrm{rad}\,\mathrm{m}^{-2}$ in the mean and  $300\, \mathrm{rad}\,\mathrm{m}^{-2}$ in the uncertainty to highlight disc features of the Galactic Faraday rotation sky.}
\end{figure*}
\begin{figure*}
\begin{subfigure}{\textwidth}
\centering
\import{\imagepathstatic}{old_faraday_wo_ff_mean.pgf}
\caption{\label{fig:faraday_mean_old}}
\end{subfigure}
\begin{subfigure}{\textwidth}
\import{\imagepath}{\run _faraday_sky_mean_diff.pgf}
\caption{\label{fig:faraday_diff}}
\end{subfigure}
\caption{\label{fig:results_old}  
\Cref{fig:faraday_mean_old} shows the mean of the Galactic Faraday rotation sky as inferred by \citetalias{2020A&A...633A.150H} using the the same saturation scale as in \cref{fig:results_250_mean}. \Cref{fig:faraday_diff} shows the difference between  \cref{fig:results_250_mean} and \cref{fig:faraday_mean_old}.
}
\end{figure*}

\begin{figure*}
\begin{subfigure}{\textwidth}
\centering
\import{\imagepath}{\run _dm_mean.pgf}
\caption{\label{fig:dm}}
\end{subfigure}
\begin{subfigure}{\textwidth}
\centering
\import{\imagepath}{\run _B_par_gal_mean.pgf}
\caption{\label{fig:chi}}
\end{subfigure}
\caption{\label{fig:results_components}  
Posterior means of the components defined in Eq. \eqref{eq:model}. 
\Cref{fig:dm} shows the amplitude field $e^\rho$ and \cref{fig:chi} shows the sign field $\chi$.
}
\end{figure*}
\begin{figure*}
\import{\imagepath}{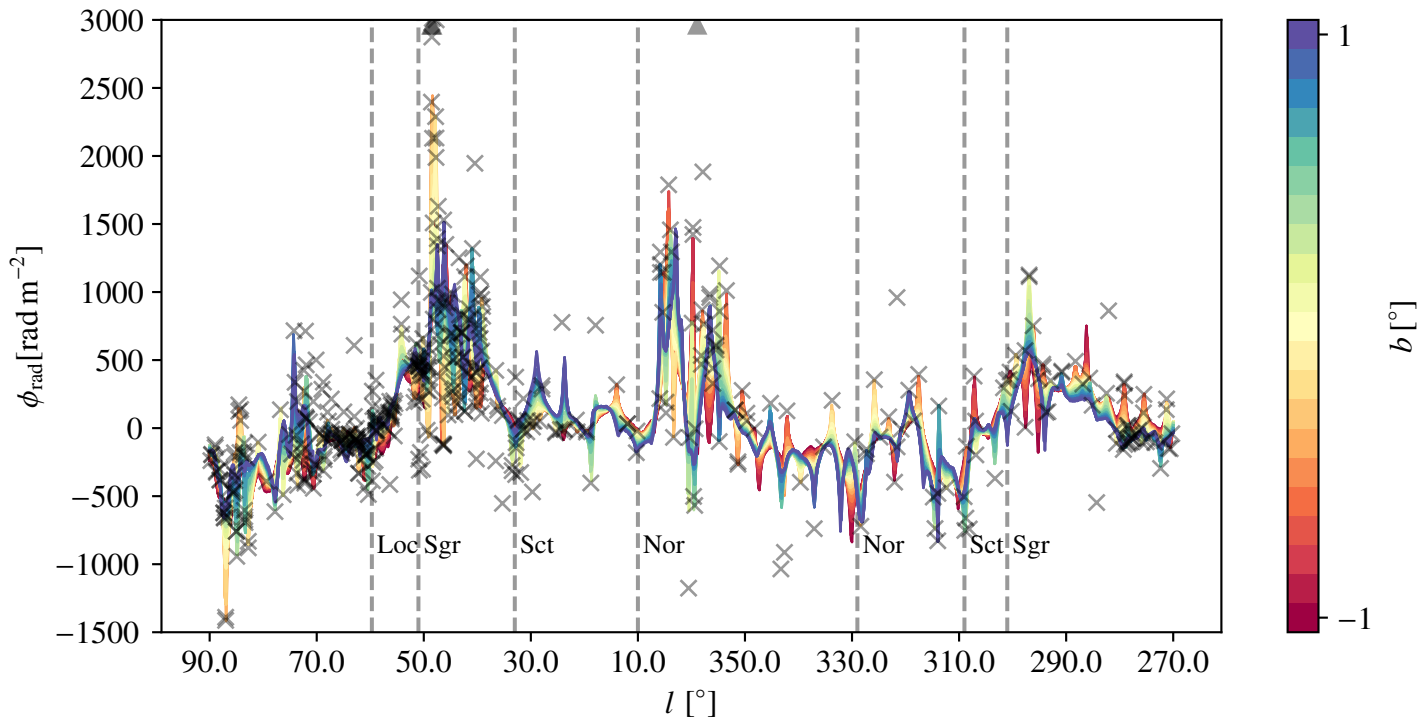}
\caption{\label{fig:slice_faraday}
A slice in longitude ($l \in (\ang{270},\ang{90})$ ) through the inner part of the Galactic disc from the Galactic Faraday rotation sky. The colored strips indicate equal latitude strips of the inferred Faraday sky within $b \in (-\ang{1},\ang{1})$. The plot also contains the data points falling in the same region indicated by grey crosses. 
Data points with  $\phi > 3000\, \mathrm{rad}\,\mathrm{m}^{-2}$ are indicated with black triangles at the upper end of the plot. 
The approximate locations of several spiral arm tangents \citep{2009A&A...499..473H, 2017AstRv..13..113V} are indicated as grey lines, specifically these are the Local (Loc), Sagittarius (Sgr), Scrutum (Sct) and Norma (Nor) arms.
Note that these positions are inferred from different tracers such as HII regions and hence do not necessarily coincide exactly with the tangent points traced by a potential excess in Faraday rotation.
The region towards $l \approx \ang{45}$ is also partly shown in \citet{2019ApJ...887L...7S}, where the strong excess in Faraday rotation in direction to the tangent of the Sagittarius arm was first noted.
}
\end{figure*}
\begin{figure}
\import{\imagepathstatic}{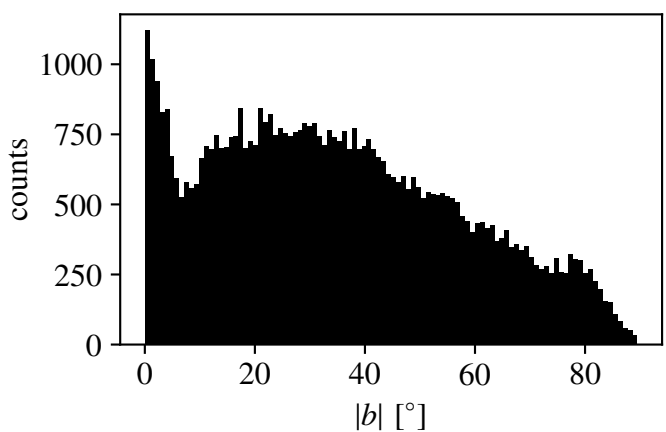}
\caption{\label{fig:lat_hist} Histogram of absolute values of the Galactic latitude positions of all data points used in this work}
\end{figure}
 
\section{Modeling the Faraday rotation sky}
\label{sec:model}

\subsection{Physics}
\label{subsec:phyics}

To provide context for the modeling of the Faraday sky, we first give a short summary on important physical concepts and general data aquisition methods for Faraday data.\par
The differential angle of rotation, $\Delta_{\psi_\lambda}$, that the polarization plane of linearly polarized light experiences between emission and detection can be described by
\begin{equation}
\label{eq:faraday_angle}
\psi_\lambda  - \psi_0 =  \Delta_{\psi_\lambda} = \mathrm{RM}\,\lambda^2,
\end{equation}
where $\lambda$ is the observational wavelength, $\psi_0$ is the intrinsic polarization angle,  $\psi_\lambda$ the angle observed at wavelength $\lambda$ and RM is the rotation measure \citep{1966MNRAS.133...67B}. 
Determining RMs has traditionally been done by observing $\psi_\lambda$ at different wavelengths, and then determining the slope of $\Delta_{\psi_\lambda}$ in $\lambda^2$ space. 
In the ideal case of a thin non-emitting plasma screen being the only source for the rotation effect, the RM at redshift zero is equal to the Faraday depth,
\begin{equation}
\label{eq:faraday_physics}
\phi(r)  = \frac{e^3}{2\pi m_e^2 c^4}\int_{r}^0 \mathrm{d}l\, n_{\mathrm{th}}(r) B_{\mathrm{\parallel}}(r),
\end{equation}
where $n_{\mathrm{th}}$ is the thermal electron density and $B_{\mathrm{\parallel}}$ is the component of the magnetic field parallel to the LOS. 
The physical constants $e, m_e$ and $c$ describe the elementary charge, the electron mass and the speed of light, respectively.
We conform with the general sign convention that RM and $\phi$ are positive for $B_{\mathrm{parallel}}$ pointing toward the observer.
Unfortunately, the relation RM $= \phi$ is generally not correct, for example in case of multiple components along the LOS.
Measurements of $\phi$ are therefore often obtained via more elaborate reconstruction techniques, such as RM synthesis \citep{2005A&A...441.1217B, 2012A&A...540A..80B}. 
In this work we will use data for $\phi$ that were determined either with a simple slope fit in $\lambda^2$ space or via RM synthesis. 
The sampling in $\lambda^2$ space is often sparse, and this sparsity is generally understood to produce artificial  feature in the Faraday spectrum \citep{2011AJ....141..191F}.
The likelihood of these being mistaken as true $\phi$-values can be minimized by careful analysis, but not completely eradicated and hence has to be considered when using the data.\par
In this work, we are interested in the Galactic component of the Faraday rotation sky $\phi_\mathrm{gal}$. 
Due to the additivity of $\phi$, one can write the equation connecting the data $d_\phi$ with $\phi_\mathrm{gal}$ as
\begin{equation}
d_\phi = \mathcal{R}\phi + n_\phi =  \mathcal{R}\left(\phi_\mathrm{gal} + \phi_\mathrm{etc} \right)  + n_\phi =  \mathcal{R}\phi_\mathrm{gal} + \widetilde{n}_\phi,  
\end{equation}
where the $\mathcal{R}$ is a projection operator connecting the sky with data space and the (Gaussian) noise term $n_\phi$ contains the known observational error. 
The non-Galactic component $\phi_\mathrm{etc}$ contains for example extra-Galactic, ionospheric contributions or unresolvable small scale structure and is absorbed together with the observational noise into an adapted noise term $\widetilde{n}_\phi$, whose covariance needs to be estimated. 
It should be emphasized at this point that, for the lack of accurate radial information connected to the RM data, our decision criterion on the discrimination between Galactic and extra-Galactic solely relies on angular correlations.
This implies that relatively close-by extra-Galactic objects such as the Magellanic clouds can end up in the Galactic map, while unresolvable small scale Galactic structures might be erased by the noise estimation.
A more accurate separation most likely would rely on detailed three dimensional modeling of the ISM and the inclusion of distance information, which by far exceeds the scope of this work. 
The noise estimation is done using the same noise estimation technique as first described by \citet{2012A&A...542A..93O}, i.e. by modeling the adapted noise standard deviations $\widetilde{\sigma}_{d_\phi}$ via
\begin{equation}
\label{eq:noise_model}
\widetilde{\sigma}^2_{d_\phi} = \eta_\phi\,\sigma^2_{d_\phi},
\end{equation} 
where $\eta_\phi$ is a parameter to be inferred via an inverse gamma model, 
\begin{equation}
\label{eq:eta_model}
\mathcal{P}\left(\eta_\phi|\alpha_\phi, \beta_\phi \right) = \frac{\beta_\phi^{\alpha_\phi}}{\Gamma(\alpha_\phi)} \eta_\phi^{-\alpha_\phi - 1} e^{-\frac{\beta_\phi}{\eta_\phi}}
\end{equation} 
as detailed by \citet{2012A&A...542A..93O} and \citetalias{2020A&A...633A.150H}.
The hyper-parameters $\alpha_\phi $ and $\beta_\phi $ steer the ability of the noise estimation to increase the noise and hence down-weights data points with large $\phi_\mathrm{etc}$ in the likelihood.
This is equivalent to setting a prior for our expectation that a data excess should be explained by the sky map or an increase of the noise.
In \citet{2012A&A...542A..93O} and \citetalias{2020A&A...633A.150H} $\alpha_\phi$ was set to a single number for the whole sky ($1$ and 2.5, respectively). 
As discussed in \citetalias{2020A&A...633A.150H} we need to adapt the hyper-parameters for the noise estimation when increasing the resolution to make up for the increase in degrees of freedom, i.e. to allow for the representation of previously unresolvable small scale structure.
With the increase of the resolution to about $1.3\cdot 10^{-2} \mathrm{deg}^2$ or $46.8\,\mathrm{arcmin}^2$ in this work, we found that setting the parameter to a restrictive value leads to over-fitting in high latitude regions (i.e. data points clearly dominated by the extra-Galactic component appearing as point sources in the sky map), while a looser choice makes it hard to represent the small scale structure in the Galactic disc.  
This indicates that we have reached a resolution where a noise model with a universally chosen $\alpha_\phi$ is not able to accurately represent the non-homogeneous statistics of the Faraday sky anymore.
We hence decided to use a position dependent noise estimation, which gets more restrictive in regions where we expect small scale structures to be mostly Galactic, e.g. towards the Galactic disc or the Magellanic clouds.  
We chose to use the logarithmic DM map of \citet{2017ApJ...835...29Y} as a proxy for this prior, as it provides a natural scaling with Galactic latitude and includes at least some important structures such as the Magellanic clouds. 
This template is then rescaled such that it reaches $1$ towards the poles (hence complying with the global value for $\alpha_\phi$ chosen in \citet{2012A&A...542A..93O}), but reaches up to $3.5$ towards the disc, implying a very restrictive prior for $\eta_\phi$ in these regions. 
%A natural extension of this model might be to dynamically couple $\alpha_\phi$ to the $\rho$ field instead of using a fixed template, as this field captures our knowledge on the amplitude of the Galactic disc in our model. 
In accordance with the previous studies, $\beta_\phi$ is then set such that the prior mode of $\eta_\phi$ is $1$ for all data points.

\begin{figure}[h!]
\centering
\begin{subfigure}{0.49\textwidth}
\import{\imagepath}{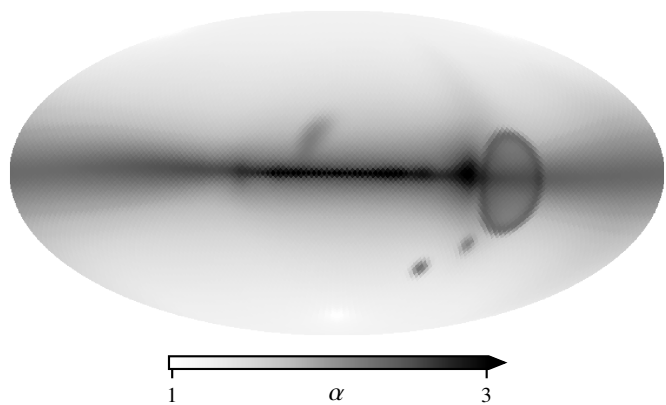}
\end{subfigure}
\caption{\label{fig:ne_alpha} 
Template for the noise estimation parameter $\alpha_\phi$ introduced in Eq. \eqref{eq:eta_model}, based on the Galactic electron model derived by \citet{2017ApJ...835...29Y}.}
\end{figure}
This noise estimation procedure also takes care of uncorrelated systematic errors as for example the $n\pi$-ambiguity, as demonstrated in \citetalias{2020A&A...633A.150H} for data points identified by \citet{2019MNRAS.487.3432M}. 
A, potentially correlated, systematic effect may come from the limited range in Faraday depth that most Faraday rotation surveys probe. 
Sources with a Faraday depth out of range are either ignored or misrepresented in the data sets, depending on the technique used to analyze the polarimetric data.
Such effects are hard to model and cannot generally be remedied by the noise estimation. Hence, we do not correct for such systematics.

\subsection{Nomenclature}
\label{subsec:nomenclature}
Before proceeding with the modeling discussion, we would like to comment on the terminology used in this paper and its predecessor \citetalias{2020A&A...633A.150H}. 
There, the authors use the term ``Faraday data'' for the data and refer to $\phi$ as Faraday depth.
Accordingly the reconstructed sky is referred to as the ``Galactic Faraday depth sky''. 
This reflects the common Bayesian ontology that clearly separates data and signal and relies on the assumption that the data are a noisy measurement of $\phi$ defined in  \cref{eq:faraday_physics}. This nomenclature is independent of the way the data were obtained or the physical geometry of the LOS observed.  
The term ``RM'' is omitted by \citetalias{2020A&A...633A.150H} and implicitly reserved for the sub set of the data set which was obtained using slope fitting techniques.
This unfortunately does not completely match the general way these terms are used in the astrophysical literature, where  sometimes the naming convention is more determined by the LOS structure. 
The label ``Faraday depth'' for \cref{eq:faraday_physics} is sometimes reserved for cases where rotation and emission are mixed and one has to deal with complex Faraday spectra (e.g. see \citealt{2018Galax...6..112V}), while ``RM'' is often used in situations that can be explained with a single non-emitting Faraday screen (e.g. see \citealt{2010ApJ...714.1170M}). 
This terminology is hard to maintain in our work as we are mixing these data sets and are agnostic to the geometry along the LOS, but on the other hand the term ``Faraday depth sky'' as used by \citetalias{2020A&A...633A.150H} might give a false impression of what is actually inferred. 
We hence decided to again label \cref{eq:faraday_physics} as ``Faraday depth'', but refer to the reconstructed sky simply as the ``Faraday rotation sky''. 

\subsection{Sky model}
\label{subsec:sky_model}
The sky model for $\phi_\mathrm{gal}$ has been well motivated by \citetalias{2020A&A...633A.150H}. 
Here, we give a very short alternative motivation, which provides connection points for additional data sets (e.g. on the Galactic  dispersion measure (DM)).
An illustrative picture arises if we assume no correlation between  $n_{\mathrm{th}}$ and $B_{\mathrm{\parallel}}$, which is reasonable for the warm ionized medium of the Galaxy \citep{2003A&A...398..845P, 2011ApJ...736...83H, 2015NewA...34...21W}, \cref{eq:faraday_physics} can be written for the Galactic component as 
\begin{align}
\label{eq:faraday_Bn_crosscorr_simple}
\phi_\mathrm{gal} \propto \mathrm{DM}\, \left\langle B_{\mathrm{\parallel}}\right\rangle_\mathrm{LOS}
\end{align}
\citep{drainebook}, where $\mathrm{DM} =\int_\mathrm{LOS} \mathrm{d}l \, n_\mathrm{th}$ and $\left\langle B_{\mathrm{\parallel}}\right\rangle_\mathrm{LOS}$ is the LOS-parallel component of the magnetic field vector averaged along the LOS. 
This equation is used to motivate the simplest model for the Faraday sky,
\begin{align}
\label{eq:model}
\phi_\mathrm{gal} = e^\rho\chi\, \mathrm{rad}\;\mathrm{m}^{-2},
\end{align}
implemented for the first time by \citetalias{2020A&A...633A.150H}. 
Here, $\rho$ and $\chi$ are Gaussian fields on the sky with unknown correlation structure that needs to be determined. 
The log-normal field $e^\rho$ is supposed to take over the role of the DM, while the sign field $\chi$ models the magnetic field average.
As already noted by \citetalias{2020A&A...633A.150H}, one cannot break the degeneracy between $e^\rho$ and $\chi$ in this model and relate the component fields to the respective physical quantities without further information and/or assumptions. 
This is attempted in a separate work \citep{HE_in_prep}. 
Note that the assumption of no correlation between $n_{\mathrm{th}}$ and $B_{\mathrm{\parallel}}$ is by no means necessary to motivate the model above as a separation of the Faraday sky into an amplitude and a sign field is of course always possible, but is only used to give an illustrative image.\par
The inference of the correlation functions has been updated in accordance with recent developments in numerical information field theory, for details see \citet{2020arXiv200205218A}. 
The new model is equally flexible in representing different power spectra, but has the advantage of more intuitive and better decoupled hyper-parameters. This inference was implemented in the newest version (v.7) of the NIFTy package \citep{NIFTY}. 
Nifty is based on Information Field Theory (IFT) and provides the user with a library of Bayesian signal inference techniques, mostly aimed at the evaluation of noisy and very high dimensional problems. 
Specifically, it makes use of the MGVI algorithm \citep{knollmueller2020metric} to approximate posterior distributions via variational inference.
It furthermore contains a toolkit to implement complicated likelihoods and signal models.
For a generic reference and introduction to IFT see \citet{2019AnP...53100127E}.

\section{Data}
\label{sec:data}

We have used a newly compiled master catalog of published Faraday data from extra-Galactic sources such as AGNs \footnote{\citet{1980A&AS...39..379T, 1981ApJS...45...97S,  1988Ap&SS.141..303B, 1992ApJ...386..143C, 1993AJ....106..444W, 1995ApJ...445..624O, 1996ApJ...458..194M, 2001ApJ...549..959G, 2001ApJ...547L.111C, 2003A&A...406..579K, 2003ApJS..145..213B,  2005MNRAS.360.1305R, 2007ApJ...663..258B, 2008A&A...487..865R, 2008ApJ...688.1029M, 2009A&A...503..409H, 2009ApJ...707..114F, 2009ApJ...702.1230T, 2010ApJ...714.1170M,
2011ApJ...728...57L, 2011ApJ...728...97V, 2011MNRAS.413..132B, 2012ApJ...755...21M, 2012ApJ...759...25M, 2015ApJ...815...49A, 2016ApJ...829..133K, 2016ApJ...821...92C, 2017MNRAS.467.1776K, 2017MNRAS.469.4034O, 2018A&A...613A..58V, 2018ApJ...865...65C, 2018MNRAS.475.1736V, 2018PASA...35...43R, 2019ApJ...887L...7S, 2019ApJ...871..215B, 2019MNRAS.485.1293S, 2019MNRAS.487.3432M, 2020MNRAS.497.3097M, 2020PASA...37...29R}
}, to be published by \citet{van_Eck_catalog}\footnote{This master catalog can be found online at https://github.com/Cameron-Van-Eck/RMTable; we used used version 0.1.8 of the catalog in this paper.}.
This catalog was assembled independently of the \citet{2012A&A...542A..93O} catalog, but contains most of the same sources plus many published since then. 
A list of the individual catalogs included can be found online with the master catalog. \par
This new master catalog is not yet complete, with papers being prioritized for inclusion based on combination of catalog size, recency, ease of data access (i.e., Vizier or machine readable tables were preferred to \LaTeX\, or PDF tables, which were preferred to images of tables). 
The version of the catalog used here includes data points from 38 papers, which encompasses nearly all of the catalogs published in the past 30 years that contain more than 30 data points.
Some papers with fewer data points were also included.  
Sources reported with multiple data points (from RM synthesis or QU-fitting) had all components included as separate entries in the catalog. 
This results in a catalog of 50,207 data points, which is a significant expansion on the 41,330 used by \cite{2020A&A...633A.150H, 2012A&A...542A..93O}.
In accordance with \citet{2011ApJ...726....4S}, the error bars of \citet{2009ApJ...702.1230T} where multiplied by $1.22$. \par
Additionally, we were provided with yet unpublished catalogs, mostly including data compiled in 2020 by the \textit{LOFAR Two-metre Sky Survey} (LoTSS) (\citet{osullivan_lotss}; 2461 data points), by the
\textit{Canadian Galactic Plane Survey} (CGPS) (\citet{van_Eck_cgps}; 2493 data points) and a small data set provided by \citet{johnstonhollit} (68 data points) which was already present in the Oppermann catalog.  
These data sets will become part of the catalog once published.
We have decided in favor of including this data, as we would like our results to represent the most up to date state of the Galactic Faraday sky as of the end of the year 2020.\par
In summary, we therefore have 55190 data points from 41 surveys available.
This number excludes 39 pulsars which were removed from the catalog, as they in general do not probe the full Galactic LOS \citep{2019MNRAS.484.3646S}. \par
We have not attempted to identify duplicate sources in the catalog; 
if a source appears in multiple catalogs, all measurements are kept.
While multiple measurements of the same source provide valuable additional information, the noise estimation technique of \citet{2012A&A...542A..93O} is strictly speaking not optimal in these cases, as these sources probe the same part of the IGM and the assumption of independence for the error estimates, which includes the extra-Galactic signal, does not hold anymore. 
The severity of this issue depends on the number of duplicates, which is hard to determine without extensive source matching. 
We can, however, give an estimate on this number by counting the number of data points located in a sky pixel containing more than one data point. 
This results in a fraction of $\approx 0.6 \%$ of the data set being identified as potential duplicates at a resolution of about $\ang{0.1}$. 
Only a small fraction of those will be actual duplicates, on the other hand close by sources might be correlated by the IGM on such scales even if they are not duplicates, as demonstrated by \citet{2010ApJ...723..476A}.
We hence assume the number of data points with extra-Galactic correlations to be on the order of $1 \%$ and therefore deem the approximation of uncorrelated noise acceptable.
In future studies of the Faraday sky, this might have to be revised. 
We assumed an a-priori noise level of $50 \%$ of $|d_\phi|$ for the 72 sources without or with zero-valued error bars, which were then subject to the same noise estimation procedure as all other data points. 
This is a rather cautious choice, but we deem the potential downsides of systematic effects introduced by `overconfident' data higher than of a potentially ignored data point due to too large error bars. 
In any case, this choice is automatically corrected to some degree by the noise estimation.
A projection of all data points, their error bars and the source density on the sky is shown in \cref{fig:faraday_data}.

\begin{figure*}
\setkeys{Gin}{draft=false}
\begin{subfigure}{0.49\textwidth}
\import{\imagepath}{\run _faraday_sky_mean_magellanic.pgf}
\caption{\label{fig:magellanic_phi}}
\end{subfigure}
\begin{subfigure}{0.49\textwidth}
\import{\imagepath}{\run _dm_mean_magellanic.pgf}
\caption{\label{fig:magellanic_dm}}
\end{subfigure}
\begin{subfigure}{0.5\textwidth}
\import{\imagepathstatic}{halpha_data_magellanic.pgf}
\caption{\label{fig:magellanic_halpha}}
\end{subfigure}
\begin{subfigure}{0.5\textwidth}
\import{\imagepathstatic}{EM_data_magellanic.pgf}
\caption{\label{fig:magellanic_em}}
\end{subfigure}
\caption{\label{fig:magellanic}  
The Magellanic Clouds in our results and as seen with different observables, respectively. \cref{fig:magellanic_phi} and \cref{fig:magellanic_dm} show the Faraday rotation (with a $20\,\mathrm{rad}\,\mathrm{m}^{-2}$ offset) and the underlying amplitude field as inferred in this work. \Cref{fig:magellanic_halpha} and \cref{fig:magellanic_em} show the clouds in $\mathrm{H}\alpha$ emission \citep{2003ApJS..146..407F, 2001PASP..113.1326G}, and in emission measure (EM) as obtained by the \textit{Planck} satellite \citep{2016A&A...594A..10P}, respectively. \cref{fig:magellanic_halpha} indicates the locations of the clouds.}
\end{figure*}

\section{Results and Discussion}
\label{sec:results}

In the following we present the results of the inference using the model introduced in \cref{sec:model} and the data presented in \cref{sec:data}.
\begin{figure}
\centering
\import{\imagepath}{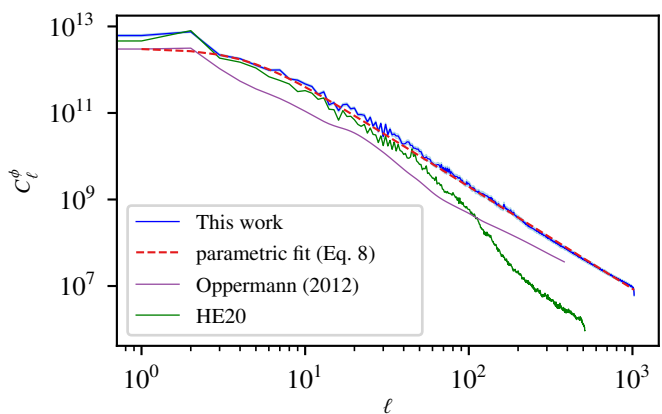}
\caption{\label{fig:power}
Power spectrum for the Faraday rotation sky derived in this work, compared with results from \citetalias{2020A&A...633A.150H} and \citet{2012A&A...542A..93O}. The red dashed line indicates a parametric fit to the results of this work using \cref{eq:power_fit}. The uncertainties of the spectrum derived in this work are indicated by lightblue lines, which are only slightly offset from the darkblue-lined mean.}
\end{figure}
\begin{figure}
%\import{\imagepath}{ \run _faraday_lat_\run _faraday_data_abs_residual.pgf}
\caption{\label{fig:lat_hist_residual} Two dimensional histogram of logarithmic noise weighted residuals and absolute Galactic latitude.}
\end{figure}

\subsection{The sky} 
\label{subsec: sky}
We show the resulting Faraday rotation sky and uncertainties in \cref{fig:results_25080} and \cref{fig:results_1500300} with different saturation scales to highlight the sky morphology at different latitude scales. 
For comparison, the previous result of \citetalias{2020A&A...633A.150H} is shown in \cref{fig:faraday_mean_old} and a difference map between this figure and \cref{fig:results_250_mean} is shown in \cref{fig:faraday_diff}. 
On large scales, the results are in good agreement with each other, while on smaller scales many more structures are discernible in the new reconstruction.
At higher latitudes, the largest deviations are discernible in the Southern sky, i.e. the region south of $-\ang{40}$ in celestial declination, which was severely under-constrained in previous inferences but has since been filled mostly by Faraday data from \citet{2019MNRAS.485.1293S}, although the region still has not reached a number density comparable to the Northern Hemisphere.
Moreover, \cref{fig:faraday_diff} demonstrates considerable differences in the Galactic disc, in parts in morphology but also in amplitude.
This can be well observed in \cref{fig:results_1500_mean}, which reveals a pronounced disc with absolute values often exceeding $1500\, \mathrm{rad\, m^{-2}}$. 
The maximum is close to $3000\, \mathrm{rad\, m^{-2}}$, which stands in strong contrast to the previous inference results of \citetalias{2020A&A...633A.150H}, which had found a maximum magnitude slightly above $1000\, \mathrm{rad\, m^{-2}}$. 
The disc is further investigated in \cref{subsubsec: inner_disc}.\par
We further show the component fields $e^\rho$ and $\chi$ in \cref{fig:results_components}. 
Both show strong similarity to the analog component fields found in \citetalias{2020A&A...633A.150H}. 
If we assume that our model in \cref{eq:model} is correct and that both maps are a proxy for the physical components of the Faraday sky, then, at least in terms of morphology, \cref{fig:dm} should trace the Galactic DM map, while \cref{fig:chi} should trace the LOS-averaged magnetic field strength.
We give some evidence for this claim in \cref{subsubsec: magellanic} and \cref{subsubsec: amplitude_correlations}, while a further more detailed investigation is to be presented by \citet{HE_in_prep}.
The $\chi$ map shows some evidence for point-like structures with opposite sign to their diffuse surroundings, which at high latitudes we cannot exclude to be remaining extra-Galactic residuals. 
This indicates that our noise estimation routine was only partially successful, and that some precautions have to be taken if the maps are used as foreground reduction template.

\subsection{Substructures}
\label{subsec: substructures}

\subsubsection{The inner disc}
\label{subsubsec: inner_disc}
The inner Galactic disc is highlighted in more detail in \cref{fig:slice_faraday}. 
Here, we demonstrate the strong variability of the disc by plotting a fine slice along Galactic longitude together with data points falling in the same region. 
The plot also contains some color coded latitude information, in order to demonstrate that within the disc the same variability is observed when moving along latitudes, including frequent sign flips. 
The plot contains the approximate directions of the Galactic spiral arm tangents. 
We note two strong excesses in amplitude of the Faraday sky, one towards the Galactic center and another one towards the tangent point of the Sagittarius arm at around $l \approx \ang{48}$ reaching almost $3000\, \mathrm{rad\, m^{-2}}$. 
The latter excess has been first reported by \citet{2019ApJ...887L...7S}, with the highest observed data point above $4000\,\mathrm{rad\, m^{-2}} $. 
Other arm locations also seem to be correlated with stronger values in $\phi$, albeit in a much less clear way. 
The Sagittarius region has also been discussed extensively by \citet{2020A&A...642A.201R}, who investigate rotation measures as spiral tracers using simulations.
In their work, the authors confirm the shape of the Sagittarius excess and attribute its sharp morphology to the geometry of the spiral arm relative to the observer.
They furthermore predict somewhat more extended but nonetheless relatively strong Faraday rotation features for the other arms and an overall increase in Faraday depth towards the Galactic center and note that these features are not visible in the results of \citet{2012A&A...542A..93O} or \citet{2020A&A...633A.150H}. 
In this work, we do find stronger disc amplitudes of the Faraday sky, but cannot confirm the above mentioned overall morphology, as our results are dominated by much smaller structures.
Assuming the consistency of our inference and that the simulations of \citet{2020A&A...642A.201R} produce a correct representation of the Milky Way, the observations can only be explained by systematic effects in the data. 
For one, the longitude region $l \in \left(\ang{150}, \ang{30}\right)$ with latitudes $|b| \in \left(\ang{3}, \ang{10}\right)$ still has under-dense data coverage, as can be seen in \cref{fig:faraday_counts} and \cref{fig:lat_hist}.
This might imply that several strong excess regions have not been noted yet as they have simply not been probed. 
Furthermore, as already noted by \citet{2020A&A...642A.201R} for the \citet{2009ApJ...702.1230T} data set and previously discussed in \cref{subsec:phyics},many Faraday rotation data sets have range limitations in $d_\phi$ and hence might misrepresent regions with exceedingly large Faraday rotation.
Modeling such systematics on the inference side, for example as an extension of \cref{eq:model} have not been attempted and potentially would require strong prior assumptions on the disc, which would be opposed to our general approach of using generic sky models in order to maintain a high flexibility.
It hence cannot be excluded that systematic effects are still present in the data, especially within the disc, and future Faraday rotation surveys are likely necessary to arrive at a complete picture.

\subsubsection{The Magellanic clouds}
\label{subsubsec: magellanic}
To investigate the correspondence of the Faraday sky with other tracers, we single out the Magellanic clouds as objects of study, due to their relatively well defined over-all morphology and small-scale structure.
The clouds are shown in \cref{fig:magellanic} as excerpts of the Faraday rotation sky, the amplitude field $e^\rho$, $\mathrm{H}\alpha$ data  \citep{2003ApJS..146..407F, 2001PASP..113.1326G} and the EM sky as obtained by \textit{Planck} \citep{2016A&A...594A..10P}, where the latter is related to the thermal electron density as $\mathrm{EM} =\int_\mathrm{LOS} \mathrm{d}l \, n^2_\mathrm{th}$ in analogy to the DM defined in Sec. \ref{subsec:sky_model}. 
A comparison between \cref{fig:magellanic_dm}  and \cref{fig:magellanic_em} reveals that the amplitude field indeed seems to trace the morphology of dense Galactic structures also on intermediate scales, as both clouds are not only present in \cref{fig:magellanic_dm}, but also are weighted correctly.
The excerpt on the Faraday sky has been offset by $20\, \mathrm{rad}\,\mathrm{m}^2$, in order to highlight the structures of the clouds.
A significant small scale correspondence of the Faraday rotation excerpt and the H$_\alpha$ is visible, which seems to indicate that even the very small scale structures are not residuals of extra-Galactic contamination, but resemble existing structures. 
Furthermore, the region between the clouds seems to indicate a coherent region of Faraday rotation between the Magellanic clouds. 
This is again consistent with an additional coherent magnetic field structure between the clouds as first reported by \citet{2008ApJ...688.1029M}, see also \citet{2017MNRAS.467.1776K}.

\subsection{Statistical properties}
\subsubsection{Power spectra}
\label{subsubsec:power_spectra}
The inference method employed in this work has seen a significant upgrade in correlation structure modeling. 
The resulting power spectrum is compared to previous results by plotting the power spectra in \cref{fig:power}.
The plot also contains the corresponding statistical uncertainties of our results.
The plot shows a significant offset of the new spectrum towards small scales compared to its predecessors. 
In order to illustrate the result, \cref{fig:power} also contains a parametric maximum a-posteriori fit to the new spectrum using the analytic formula
\begin{equation}
\label{eq:power_fit}
C^{\phi}_{\ell, \mathrm{fit}} = \frac{A}{1 + \left(\frac{\ell}{\ell_{0}}\right)^{-s}}, 
\end{equation}
which results in $s=-2.4$, $\ell_0 = 34$ and $A = 1.1\cdot\,10^{14}$.
While a similar power law slope of $\approx\,-2.17$ was found by \citet{2012A&A...542A..93O}, the power spectra on small scales were at least somewhat suppressed in \citetalias{2020A&A...633A.150H}. 
This is consistent with the new small-scale structures visible in \cref{fig:results_250_mean} as compared to \cref{fig:faraday_mean_old}. 

\begin{figure*}
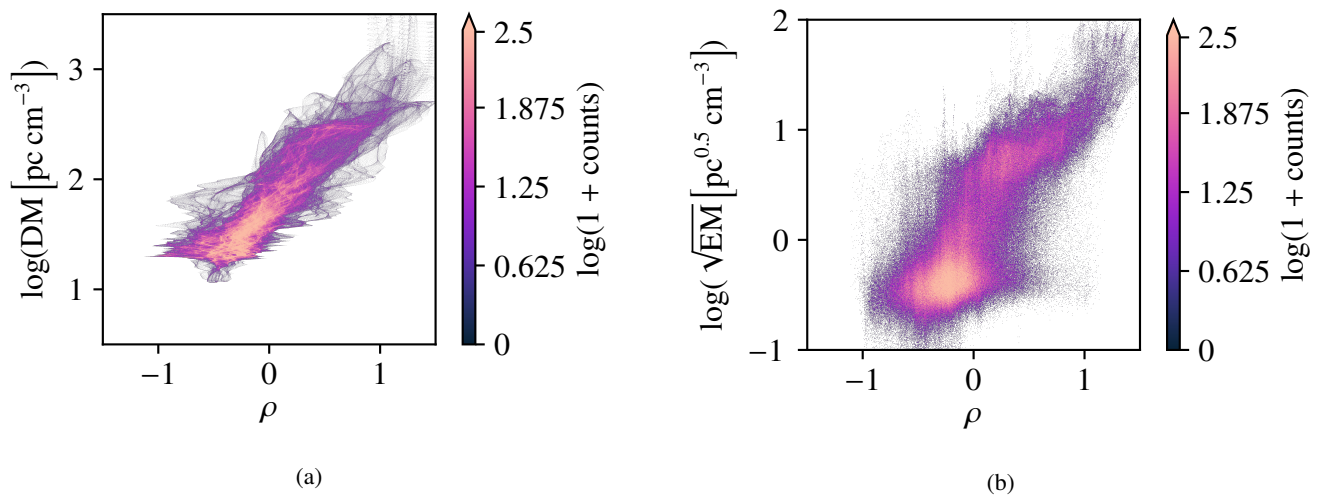

\centering
\begin{subfigure}{0.49\textwidth}
\centering
\import{\imagepath}{\run _dm_mean_yml_data_yml.pgf}
\caption{\label{fig:scatter_dm}}
\end{subfigure}
\begin{subfigure}{0.49\textwidth}
\centering
\import{\imagepath}{\run _dm_mean_freefree_data_em.pgf}
\caption{\label{fig:scatter_em}}
\end{subfigure}
\caption{\label{fig:scatter}
Histogram of sky pixels illustrating the correlation between the log amplitude field $\rho$ and the logarithmic DM (\cref{fig:scatter_dm}) and logarithmic square root EM (\cref{fig:scatter_em}) skies, respectively. The former is calculated from a Galactic thermal electron model \citep{2017ApJ...835...29Y}, while the latter was inferred by the \textit{Planck} survey from extinction-free microwave data \citep{2016A&A...594A..10P}.. }
\end{figure*}

\subsubsection{Latitude correlations}
\label{subsubsection:latitude_correlations}

An important consistency test for the convergence of our inference and a successful separation of the Galactic component of the Faraday sky is to check if there are any latitude correlations remaining in the residuals $d_\phi - \mathcal{R}\phi_{\mathrm{gal}}$. 
Since a data point can be either fitted by adjusting the sky or by increasing the noise estimate, it is best to compare the noise weighted residual $\widetilde{\sigma}_{d_\phi}^{-1}|d_\phi-\mathcal{R}\phi_\mathrm{gal}|$ as a dimensionless tracer of such correlations.
The noise $\widetilde{\sigma}_{d_\phi}$ in this case is the standard deviation as defined in \cref{eq:noise_model}.
This is investigated in \cref{fig:lat_hist_residual}, which shows a two-dimensional histogram between the logarithmic noise-weighted residuals and the absolute value of the Galactic latitude. 
For latitudes exceeding $\ang{3}$, no notable correlation with latitude is discernible. 
This is not true for the scatter of the residual, which is strong in the disc and smaller at higher latitudes.
Such an increase is however easily explained when viewing the distribution of data points along Galactic latitude as depicted in the histogram \cref{fig:lat_hist}. 
Another peculiar feature of \cref{fig:lat_hist_residual} is the relatively sharp decrease of the count density for log residuals of $0$. This is an imprint of the distribution of our uncertainties in our Faraday data catalog, which sees a relatively hard cut towards large error-bars.

\subsubsection{Amplitude correlations}
\label{subsubsec: amplitude_correlations}

The inferences conducted in \citetalias{2020A&A...633A.150H} and \citep{HE_in_prep} make use of the EM sky as a proxy for the amplitude of the Faraday sky. 
In order to test the extent to which this assumption is viable, we investigate the correlation between the amplitude field (shown in \cref{fig:dm}) and both the DM sky and the square root of the EM sky, \cref{fig:scatter}.
The square root is motivated by the quadratic dependence of the EM on the thermal electron density, in contrast to the linear dependence of DM and RM. 
The latter was inferred from cosmic microwave data by the \textit{Planck} survey \citep{2016A&A...594A..10P}, while the former is calculated from a parametric model of the Galactic thermal electron density \citep{2017ApJ...835...29Y} based on pulsar DM data. 
Although the EM map of \textit{Planck} is much more detailed and the DM map relies on very little data and strong modeling
assumptions, we note much clearer correlations of the amplitude field with the DM map than with the $\sqrt{\mathrm{EM}}$ map, aligning with both the model and the physical discussion presented in \cref{subsec:sky_model}. 
This is confirmed by Spearman-rank cross-correlation coefficients of $0.88$ for the log DM map and $0.72$ for the log square root EM map.
In the case of the DM map some of the missing correlation to the amplitude map can be explained by the fact that the \citet{2017ApJ...835...29Y} electron model lacks small-scale structure as it relies on sparse data.
The differences to the square root EM map, which itself contains a lot of small scale structure, are best explained by the missing volume filling factor that needs to be considered in order to equate the EM and DM skies.
\citet{2008PASA...25..184G} show that this factor is variable over the sky and ranges between 0.04 in the mid-plane to about 0.3 at larger distances from the Galactic disc.
As demonstrated by \citetalias{2020A&A...633A.150H}, the EM sky is nonetheless a valuable source of information on the Galactic electron density profile. 
It is however necessary to model these systematic volume effects by introducing additional degrees of freedom, as implemented phenomenologically by \citetalias{2020A&A...633A.150H}.
An improved version of this model which takes the above discussion into account will be given by \citet{HE_in_prep}.

\section{Summary and Conclusion}
\label{sec:conclusion}
We have reconstructed the Galactic Faraday rotation sky using a newly-available Faraday rotation data set, comprising the catalog compiled by \citep{van_Eck_catalog} and including additional catalogs provided by \citet{osullivan_lotss, van_Eck_cgps, johnstonhollit}. 
This work is an update to the reconstructions by \citet{2012A&A...542A..93O} and \citetalias{2020A&A...633A.150H}.
Additionally to the new data set, we have made technical advancements by employing a new correlation structure model originally developed by \citet{2020arXiv200205218A} and providing our results with an increased angular resolution. 
We have found significant updates in contrast to older results, such as an enhanced Galactic disc with Faraday rotation amplitudes near $3000\,\mathrm{rad\,m^{-2}}$ and drastically improved small-scale structures in the sky map. 
Apart from a morphological examination of the full sky and certain excerpts such as the Galactic disc and the Magellanic clouds, we have further examined the statistical properties of our results by investigating the power spectrum of the resulting map and cross correlations of the map and its components with other ISM tracers. 
We have found further motivation to the attempt begun by \citetalias{2020A&A...633A.150H} to introduce additional tracers for more holistic Galactic sky inferences. 
Our results can be used as a foreground reduction template for extra-Galactic Faraday data, however one should note that some extra-Galactic residuals might still be present in the posterior mean maps.
An ideal foreground reduction pipeline should hence mirror the analysis of \citet{2015A&A...575A.118O}, i.e. by performing a joint analysis of the extra-Galactic, systematic and Galactic components to the variance. 
If such a procedure is deemed too expensive to implement, at least some smoothing should be applied to the maps, to lessen the effect of potential extra-Galactic residuals, especially at higher latitudes.
Our analysis has furthermore demonstrated that future Faraday rotation sky inferences might strongly benefit from future surveys that remedy certain shortcomings of the existing data sets such as
\begin{itemize}
\item the still sparse data in parts of the Galactic disc, the regions above and below the disc and the southern celestial, sky (although the situation there has improved considerably in recent years, mainly due to the results of \citet{2019MNRAS.485.1293S}) and
\item the systematic biases introduced by several surveys due to their limited Faraday depth range, which might mean that several strong Faraday rotating regions are still undetected.
\end{itemize}
These issues indicate a strong potential for the upcoming polarimetric surveys of ASKAP, MeerKAT, LOFAR, MWA, VLASS, and SKA  in increasing our knowledge on the Galactic Faraday rotation sky, as they will provide a vast catalog of new Faraday data \citep{2020Galax...8...53H}. 
This information will, in turn, provide a pivotal role in constraining the Galactic magnetic field and is an important input for upcoming reconstructions of the magnetized interstellar medium. 

\section*{Acknowledgment}
The results in this publication have been derived using the
NIFTy package (https://gitlab.mpcdf.mpg.de/ift/NIFTy, \citet{NIFTY}). 
Some of the images were produced using the CMasher library (https://github.com/1313e/CMasher, \citet{2020JOSS....5.2004V}). 
We would like to thank Jennifer West and the CIRADA team providing the cutout server.
SH would like to thank Philipp Frank, Philipp Arras, Martin Reinecke, Jakob Knollm\"uller, Reimar Leike, and the rest of the IFT team for valuable discussions and their continuous work on the NIFTy package. 
The noise estimation template in Fig. \ref{fig:ne_alpha} was produced using the pygedm package (https://pypi.org/project/pygedm/, \citep{2017ApJ...835...29Y}).
We thank Rainer Beck for his careful reading and valuable comments as the Max Planck Institute for Radio Astronomy (MPIfR) internal referee.\\
SH and MH acknowledge funding from the European Research Council (ERC) under the European Union's Horizon 2020 research and innovation programme (grant agreement No 772663).
The Australia Telescope is funded by the Commonwealth of Australia for operation as a National Facility managed by CSIRO.
JMS acknowledges the support of the Natural Sciences and Engineering Research Council of Canada (NSERC), 2019-04848.
JB acknowledges the support of the Natural Sciences and Engineering Research Council of Canada and the National Research Council of Canada. 
Basic research in radio astronomy at the US Naval Research Laboratory is supported by 6.1 Base funding.
CJR acknowledges financial support from the ERC Starting Grant ``DRANOEL'', number 714245. 
C.L.H.H. acknowledges the support of the NAOJ Fellowship and JSPS KAKENHI grants 18K13586 and 20K14527.
The Paul G. Allen Family Foundation, the US Naval Observatory and the US National Science Foundation grants AST-0321309, AST-0540690 and AST-0838268 have contributed to the ATA project.
This paper has made use of the S-PASS/ATCA RM catalog \cite{2019MNRAS.485.1293S}.
The Dunlap Institute is funded through an endowment established by the David Dunlap family and the University of Toronto. C.V.E. and B.M.G. acknowledge the support of the Natural Sciences and Engineering Research Council of Canada (NSERC) through grant RGPIN-2015-05948, of the Canada Research Chairs program, and of the Canada Foundation for Innovation 2017 Innovation Fund through Project 35999.
LOFAR \citep{2013A&A...556A...2V} is the Low Frequency Array designed and constructed by ASTRON. It has observing, data processing, and data storage facilities in several countries, which are owned by various parties (each with their own funding sources), and which are collectively operated by the ILT foundation under a joint scientific policy. The ILT resources have benefited from the following recent major funding sources: CNRS-INSU, Observatoire de Paris and Universit\'e d'Orl{\'e}ans, France; BMBF, MIWF-NRW, MPG, Germany; Science Foundation Ireland (SFI), Department of Business, Enterprise and Innovation (DBEI), Ireland; NWO, The Netherlands; The Science and Technology Facilities Council, UK; Ministry of Science and Higher Education, Poland; The Istituto Nazionale di Astrofisica (INAF), Italy.
This research made use of the Dutch national e-infrastructure with support of the SURF Cooperative (e-infra 180169) and the LOFAR e-infra group. The J\"ulich LOFAR Long Term Archive and the German LOFAR network are both coordinated and operated by the J\"ulich Supercomputing Centre (JSC), and computing resources on the supercomputer JUWELS at JSC were provided by the Gauss Centre for Supercomputing e.V. (grant CHTB00) through the John von Neumann Institute for Computing (NIC).
This research made use of the University of Hertfordshire high-performance computing facility and the LOFAR-UK computing facility located at the University of Hertfordshire and supported by STFC [ST/P000096/1], and of the Italian LOFAR IT computing infrastructure supported and operated by INAF, and by the Physics Department of Turin university (under an agreement with Consorzio Interuniversitario per la Fisica Spaziale) at the C3S Supercomputing Centre, Italy.

\bibliographystyle{aa}
\bibliography{lib}

\begin{thebibliography}{87}
\expandafter\ifx\csname natexlab\endcsname\relax\def\natexlab#1{#1}\fi

\bibitem[{{Akahori} \& {Ryu}(2010)}]{2010ApJ...723..476A}
{Akahori}, T. \& {Ryu}, D. 2010, \apj, 723, 476

\bibitem[{{Anderson} {et~al.}(2015){Anderson}, {Gaensler}, {Feain}, \&
  {Franzen}}]{2015ApJ...815...49A}
{Anderson}, C.~S., {Gaensler}, B.~M., {Feain}, I.~J., \& {Franzen}, T.~M.~O.
  2015, \apj, 815, 49

\bibitem[{{Arras} {et~al.}(2020){Arras}, {Frank}, {Haim}, {Knollm{\"u}ller},
  {Leike}, {Reinecke}, \& {En{\ss}lin}}]{2020arXiv200205218A}
{Arras}, P., {Frank}, P., {Haim}, P., {et~al.} 2020, arXiv e-prints,
  arXiv:2002.05218

\bibitem[{{Battye} {et~al.}(2011){Battye}, {Browne}, {Peel}, {Jackson}, \&
  {Dickinson}}]{2011MNRAS.413..132B}
{Battye}, R.~A., {Browne}, I.~W.~A., {Peel}, M.~W., {Jackson}, N.~J., \&
  {Dickinson}, C. 2011, \mnras, 413, 132

\bibitem[{{Beck}(2015)}]{2015A&ARv..24....4B}
{Beck}, R. 2015, \aapr, 24, 4

\bibitem[{{Bell} \& {En{\ss}lin}(2012)}]{2012A&A...540A..80B}
{Bell}, M.~R. \& {En{\ss}lin}, T.~A. 2012, \aap, 540, A80

\bibitem[{{Betti} {et~al.}(2019){Betti}, {Hill}, {Mao}, {Gaensler}, {Lockman},
  {McClure-Griffiths}, \& {Benjamin}}]{2019ApJ...871..215B}
{Betti}, S.~K., {Hill}, A.~S., {Mao}, S.~A., {et~al.} 2019, \apj, 871, 215

\bibitem[{{Brentjens} \& {de Bruyn}(2005)}]{2005A&A...441.1217B}
{Brentjens}, M.~A. \& {de Bruyn}, A.~G. 2005, \aap, 441, 1217

\bibitem[{{Broten} {et~al.}(1988){Broten}, {MacLeod}, \&
  {Vallee}}]{1988Ap&SS.141..303B}
{Broten}, N.~W., {MacLeod}, J.~M., \& {Vallee}, J.~P. 1988, \apss, 141, 303

\bibitem[{{Brown} {et~al.}(2007){Brown}, {Haverkorn}, {Gaensler}, {Taylor},
  {Bizunok}, {McClure-Griffiths}, {Dickey}, \& {Green}}]{2007ApJ...663..258B}
{Brown}, J.~C., {Haverkorn}, M., {Gaensler}, B.~M., {et~al.} 2007, \apj, 663,
  258

\bibitem[{{Brown} {et~al.}(2003){Brown}, {Taylor}, \&
  {Jackel}}]{2003ApJS..145..213B}
{Brown}, J.~C., {Taylor}, A.~R., \& {Jackel}, B.~J. 2003, \apjs, 145, 213

\bibitem[{{Burn}(1966)}]{1966MNRAS.133...67B}
{Burn}, B.~J. 1966, \mnras, 133, 67

\bibitem[{{Clarke} {et~al.}(2001){Clarke}, {Kronberg}, \&
  {B{\"o}hringer}}]{2001ApJ...547L.111C}
{Clarke}, T.~E., {Kronberg}, P.~P., \& {B{\"o}hringer}, H. 2001, \apjl, 547,
  L111

\bibitem[{{Clegg} {et~al.}(1992){Clegg}, {Cordes}, {Simonetti}, \&
  {Kulkarni}}]{1992ApJ...386..143C}
{Clegg}, A.~W., {Cordes}, J.~M., {Simonetti}, J.~M., \& {Kulkarni}, S.~R. 1992,
  \apj, 386, 143

\bibitem[{{Cooper} \& {Price}(1962)}]{1962Natur.195.1084C}
{Cooper}, B.~F.~C. \& {Price}, R.~M. 1962, \nat, 195, 1084

\bibitem[{{Costa} \& {Spangler}(2018)}]{2018ApJ...865...65C}
{Costa}, A.~H. \& {Spangler}, S.~R. 2018, \apj, 865, 65

\bibitem[{{Costa} {et~al.}(2016){Costa}, {Spangler}, {Sink}, {Brown}, \&
  {Mao}}]{2016ApJ...821...92C}
{Costa}, A.~H., {Spangler}, S.~R., {Sink}, J.~R., {Brown}, S., \& {Mao}, S.~A.
  2016, \apj, 821, 92

\bibitem[{{Dineen} \& {Coles}(2005)}]{2005MNRAS.362..403D}
{Dineen}, P. \& {Coles}, P. 2005, \mnras, 362, 403

\bibitem[{Draine(2010)}]{drainebook}
Draine, B.~T. 2010, {Physics of the interstellar and intergalactic medium},
  Princeton series in astrophysics (Princeton, NJ: Princeton University Press)

\bibitem[{{Durrer} \& {Neronov}(2013)}]{2013A&ARv..21...62D}
{Durrer}, R. \& {Neronov}, A. 2013, \aapr, 21, 62

\bibitem[{{En{\ss}lin}(2019)}]{2019AnP...53100127E}
{En{\ss}lin}, T.~A. 2019, Annalen der Physik, 531, 1800127

\bibitem[{{Farnsworth} {et~al.}(2011){Farnsworth}, {Rudnick}, \&
  {Brown}}]{2011AJ....141..191F}
{Farnsworth}, D., {Rudnick}, L., \& {Brown}, S. 2011, \aj, 141, 191

\bibitem[{{Feain} {et~al.}(2009){Feain}, {Ekers}, {Murphy}, {Gaensler},
  {Macquart}, {Norris}, {Cornwell}, {Johnston-Hollitt}, {Ott}, \&
  {Middelberg}}]{2009ApJ...707..114F}
{Feain}, I.~J., {Ekers}, R.~D., {Murphy}, T., {et~al.} 2009, \apj, 707, 114

\bibitem[{{Finkbeiner}(2003)}]{2003ApJS..146..407F}
{Finkbeiner}, D.~P. 2003, \apjs, 146, 407

\bibitem[{{Frick} {et~al.}(2001){Frick}, {Stepanov}, {Shukurov}, \&
  {Sokoloff}}]{2001MNRAS.325..649F}
{Frick}, P., {Stepanov}, R., {Shukurov}, A., \& {Sokoloff}, D. 2001, \mnras,
  325, 649

\bibitem[{{Gaensler} {et~al.}(2001){Gaensler}, {Dickey}, {McClure-Griffiths},
  {Green}, {Wieringa}, \& {Haynes}}]{2001ApJ...549..959G}
{Gaensler}, B.~M., {Dickey}, J.~M., {McClure-Griffiths}, N.~M., {et~al.} 2001,
  \apj, 549, 959

\bibitem[{{Gaensler} {et~al.}(2008){Gaensler}, {Madsen}, {Chatterjee}, \&
  {Mao}}]{2008PASA...25..184G}
{Gaensler}, B.~M., {Madsen}, G.~J., {Chatterjee}, S., \& {Mao}, S.~A. 2008,
  \pasa, 25, 184

\bibitem[{{Gardner} \& {Davies}(1966)}]{1966AuJPh..19..129G}
{Gardner}, F.~F. \& {Davies}, R.~D. 1966, Australian Journal of Physics, 19,
  129

\bibitem[{{Gaustad} {et~al.}(2001){Gaustad}, {McCullough}, {Rosing}, \& {Van
  Buren}}]{2001PASP..113.1326G}
{Gaustad}, J.~E., {McCullough}, P.~R., {Rosing}, W., \& {Van Buren}, D. 2001,
  \pasp, 113, 1326

\bibitem[{{Harvey-Smith} {et~al.}(2011){Harvey-Smith}, {Madsen}, \&
  {Gaensler}}]{2011ApJ...736...83H}
{Harvey-Smith}, L., {Madsen}, G.~J., \& {Gaensler}, B.~M. 2011, \apj, 736, 83

\bibitem[{{Haverkorn}(2015)}]{2015ASSL..407..483H}
{Haverkorn}, M. 2015, {Magnetic Fields in the Milky Way}, ed. A.~{Lazarian},
  E.~M. {de Gouveia Dal Pino}, \& C.~{Melioli}, Vol. 407, 483

\bibitem[{{Heald} {et~al.}(2009){Heald}, {Braun}, \&
  {Edmonds}}]{2009A&A...503..409H}
{Heald}, G., {Braun}, R., \& {Edmonds}, R. 2009, \aap, 503, 409

\bibitem[{{Heald} {et~al.}(2020){Heald}, {Mao}, {Vacca}, {Akahori},
  {Damas-Segovia}, {Gaensler}, {Hoeft}, {Agudo}, {Basu}, {Beck}, {Birkinshaw},
  {Bonafede}, {Bourke}, {Bracco}, {Carretti}, {Feretti}, {Girart}, {Govoni},
  {Green}, {Han}, {Haverkorn}, {Horellou}, {Johnston-Hollitt}, {Kothes},
  {Landecker}, {Nikiel-Wroczy{\'n}ski}, {O'Sullivan}, {Padovani}, {Poidevin},
  {Pratley}, {Regis}, {Riseley}, {Robishaw}, {Rudnick}, {Sobey}, {Stil}, {Sun},
  {Sur}, {Taylor}, {Thomson}, {Van Eck}, {Vazza}, {West}, \& {SKA Magnetism
  Science Working Group}}]{2020Galax...8...53H}
{Heald}, G., {Mao}, S., {Vacca}, V., {et~al.} 2020, Galaxies, 8, 53

\bibitem[{{Hou} {et~al.}(2009){Hou}, {Han}, \& {Shi}}]{2009A&A...499..473H}
{Hou}, L.~G., {Han}, J.~L., \& {Shi}, W.~B. 2009, \aap, 499, 473

\bibitem[{{Hutschenreuter} \& {En{\ss}lin}(2020)}]{2020A&A...633A.150H}
{Hutschenreuter}, S. \& {En{\ss}lin}, T.~A. 2020, \aap, 633, A150

\bibitem[{{Hutschenreuter} \& {Ensslin}(in prep.)}]{HE_in_prep}
{Hutschenreuter}, S. \& {Ensslin}, T.~A. in prep.

\bibitem[{{Johnston-Hollit et al.}(in prep.)}]{johnstonhollit}
{Johnston-Hollit et al.} in prep.

\bibitem[{{Johnston-Hollitt} {et~al.}(2004){Johnston-Hollitt}, {Hollitt}, \&
  {Ekers}}]{2004mim..proc...13J}
{Johnston-Hollitt}, M., {Hollitt}, C.~P., \& {Ekers}, R.~D. 2004, in The
  Magnetized Interstellar Medium, ed. B.~{Uyaniker}, W.~{Reich}, \&
  R.~{Wielebinski}, 13--18

\bibitem[{{Kaczmarek} {et~al.}(2017){Kaczmarek}, {Purcell}, {Gaensler},
  {McClure-Griffiths}, \& {Stevens}}]{2017MNRAS.467.1776K}
{Kaczmarek}, J.~F., {Purcell}, C.~R., {Gaensler}, B.~M., {McClure-Griffiths},
  N.~M., \& {Stevens}, J. 2017, \mnras, 467, 1776

\bibitem[{{Kim} {et~al.}(2016){Kim}, {Lilly}, {Miniati}, {Bernet}, {Beck},
  {O'Sullivan}, \& {Gaensler}}]{2016ApJ...829..133K}
{Kim}, K.~S., {Lilly}, S.~J., {Miniati}, F., {et~al.} 2016, \apj, 829, 133

\bibitem[{{Klein} {et~al.}(2003){Klein}, {Mack}, {Gregorini}, \&
  {Vigotti}}]{2003A&A...406..579K}
{Klein}, U., {Mack}, K.~H., {Gregorini}, L., \& {Vigotti}, M. 2003, \aap, 406,
  579

\bibitem[{Knollmüller \& Enßlin(2020)}]{knollmueller2020metric}
Knollmüller, J. \& Enßlin, T.~A. 2020 [\eprint[arXiv]{1901.11033}]

\bibitem[{{Law} {et~al.}(2011){Law}, {Gaensler}, {Bower}, {Backer},
  {Bauermeister}, {Croft}, {Forster}, {Gutierrez-Kraybill}, {Harvey-Smith},
  {Heiles}, {Hull}, {Keating}, {MacMahon}, {Whysong}, {Williams}, \&
  {Wright}}]{2011ApJ...728...57L}
{Law}, C.~J., {Gaensler}, B.~M., {Bower}, G.~C., {et~al.} 2011, \apj, 728, 57

\bibitem[{{Ma} {et~al.}(2020){Ma}, {Mao}, {Ordog}, \&
  {Brown}}]{2020MNRAS.497.3097M}
{Ma}, Y.~K., {Mao}, S.~A., {Ordog}, A., \& {Brown}, J.~C. 2020, \mnras, 497,
  3097

\bibitem[{{Ma} {et~al.}(2019){Ma}, {Mao}, {Stil}, {Basu}, {West}, {Heiles},
  {Hill}, \& {Betti}}]{2019MNRAS.487.3432M}
{Ma}, Y.~K., {Mao}, S.~A., {Stil}, J., {et~al.} 2019, \mnras, 487, 3432

\bibitem[{{Mao} {et~al.}(2010){Mao}, {Gaensler}, {Haverkorn}, {Zweibel},
  {Madsen}, {McClure-Griffiths}, {Shukurov}, \&
  {Kronberg}}]{2010ApJ...714.1170M}
{Mao}, S.~A., {Gaensler}, B.~M., {Haverkorn}, M., {et~al.} 2010, \apj, 714,
  1170

\bibitem[{{Mao} {et~al.}(2008){Mao}, {Gaensler}, {Stanimirovi{\'c}},
  {Haverkorn}, {McClure-Griffiths}, {Staveley-Smith}, \&
  {Dickey}}]{2008ApJ...688.1029M}
{Mao}, S.~A., {Gaensler}, B.~M., {Stanimirovi{\'c}}, S., {et~al.} 2008, \apj,
  688, 1029

\bibitem[{{Mao} {et~al.}(2012{\natexlab{a}}){Mao}, {McClure-Griffiths},
  {Gaensler}, {Brown}, {van Eck}, {Haverkorn}, {Kronberg}, {Stil}, {Shukurov},
  \& {Taylor}}]{2012ApJ...755...21M}
{Mao}, S.~A., {McClure-Griffiths}, N.~M., {Gaensler}, B.~M., {et~al.}
  2012{\natexlab{a}}, \apj, 755, 21

\bibitem[{{Mao} {et~al.}(2012{\natexlab{b}}){Mao}, {McClure-Griffiths},
  {Gaensler}, {Haverkorn}, {Beck}, {McConnell}, {Wolleben}, {Stanimirovi{\'c}},
  {Dickey}, \& {Staveley-Smith}}]{2012ApJ...759...25M}
{Mao}, S.~A., {McClure-Griffiths}, N.~M., {Gaensler}, B.~M., {et~al.}
  2012{\natexlab{b}}, \apj, 759, 25

\bibitem[{{Minter} \& {Spangler}(1996)}]{1996ApJ...458..194M}
{Minter}, A.~H. \& {Spangler}, S.~R. 1996, \apj, 458, 194

\bibitem[{{Oppermann} {et~al.}(2015){Oppermann}, {Junklewitz}, {Greiner},
  {En{\ss}lin}, {Akahori}, {Carretti}, {Gaensler}, {Goobar}, {Harvey-Smith},
  {Johnston-Hollitt}, {Pratley}, {Schnitzeler}, {Stil}, \&
  {Vacca}}]{2015A&A...575A.118O}
{Oppermann}, N., {Junklewitz}, H., {Greiner}, M., {et~al.} 2015, \aap, 575,
  A118

\bibitem[{{Oppermann} {et~al.}(2012){Oppermann}, {Junklewitz}, {Robbers},
  {Bell}, {En{\ss}lin}, {Bonafede}, {Braun}, {Brown}, {Clarke}, {Feain},
  {Gaensler}, {Hammond}, {Harvey-Smith}, {Heald}, {Johnston-Hollitt}, {Klein},
  {Kronberg}, {Mao}, {McClure-Griffiths}, {O'Sullivan}, {Pratley}, {Robishaw},
  {Roy}, {Schnitzeler}, {Sotomayor-Beltran}, {Stevens}, {Stil}, {Sunstrum},
  {Tanna}, {Taylor}, \& {Van Eck}}]{2012A&A...542A..93O}
{Oppermann}, N., {Junklewitz}, H., {Robbers}, G., {et~al.} 2012, \aap, 542, A93

\bibitem[{{Oren} \& {Wolfe}(1995)}]{1995ApJ...445..624O}
{Oren}, A.~L. \& {Wolfe}, A.~M. 1995, \apj, 445, 624

\bibitem[{{O'Sullivan} {et~al.}(2017){O'Sullivan}, {Purcell}, {Anderson},
  {Farnes}, {Sun}, \& {Gaensler}}]{2017MNRAS.469.4034O}
{O'Sullivan}, S.~P., {Purcell}, C.~R., {Anderson}, C.~S., {et~al.} 2017,
  \mnras, 469, 4034

\bibitem[{{O'Sullivan et al.}(in prep.)}]{osullivan_lotss}
{O'Sullivan et al.} in prep.

\bibitem[{{Passot} \& {V{\'a}zquez-Semadeni}(2003)}]{2003A&A...398..845P}
{Passot}, T. \& {V{\'a}zquez-Semadeni}, E. 2003, \aap, 398, 845

\bibitem[{{Planck Collaboration} {et~al.}(2016){Planck Collaboration}, {Adam},
  {Ade}, {Aghanim}, {Alves}, {Arnaud}, {Ashdown}, {Aumont}, {Baccigalupi},
  {Banday}, \& et~al.}]{2016A&A...594A..10P}
{Planck Collaboration}, {Adam}, R., {Ade}, P.~A.~R., {et~al.} 2016, \aap, 594,
  A10

\bibitem[{{Pshirkov} {et~al.}(2011){Pshirkov}, {Tinyakov}, {Kronberg}, \&
  {Newton-McGee}}]{2011ApJ...738..192P}
{Pshirkov}, M.~S., {Tinyakov}, P.~G., {Kronberg}, P.~P., \& {Newton-McGee},
  K.~J. 2011, \apj, 738, 192

\bibitem[{{Reissl} {et~al.}(2020){Reissl}, {Stil}, {Chen}, {Tre{\ss}},
  {Sormani}, {Smith}, {Klessen}, {Buick}, {Glover}, {Shanahan}, {Lemmer},
  {Soler}, {Beuther}, {Urquhart}, {Anderson}, {Menten}, {Brunthaler}, {Ragan},
  \& {Rugel}}]{2020A&A...642A.201R}
{Reissl}, S., {Stil}, J.~M., {Chen}, E., {et~al.} 2020, \aap, 642, A201

\bibitem[{{Riseley} {et~al.}(2020){Riseley}, {Galvin}, {Sobey}, {Vernstrom},
  {White}, {Zhang}, {Gaensler}, {Heald}, {Anderson}, {Franzen}, {Hancock},
  {Hurley-Walker}, {Lenc}, \& {Van Eck}}]{2020PASA...37...29R}
{Riseley}, C.~J., {Galvin}, T.~J., {Sobey}, C., {et~al.} 2020, \pasa, 37, e029

\bibitem[{{Riseley} {et~al.}(2018){Riseley}, {Lenc}, {Van Eck}, {Heald},
  {Gaensler}, {Anderson}, {Hancock}, {Hurley-Walker}, {Sridhar}, \&
  {White}}]{2018PASA...35...43R}
{Riseley}, C.~J., {Lenc}, E., {Van Eck}, C.~L., {et~al.} 2018, \pasa, 35, 43

\bibitem[{{Rossetti} {et~al.}(2008){Rossetti}, {Dallacasa}, {Fanti}, {Fanti},
  \& {Mack}}]{2008A&A...487..865R}
{Rossetti}, A., {Dallacasa}, D., {Fanti}, C., {Fanti}, R., \& {Mack}, K.~H.
  2008, \aap, 487, 865

\bibitem[{{Roy} {et~al.}(2005){Roy}, {Rao}, \&
  {Subrahmanyan}}]{2005MNRAS.360.1305R}
{Roy}, S., {Rao}, A.~P., \& {Subrahmanyan}, R. 2005, \mnras, 360, 1305

\bibitem[{{Schnitzeler} {et~al.}(2019){Schnitzeler}, {Carretti}, {Wieringa},
  {Gaensler}, {Haverkorn}, \& {Poppi}}]{2019MNRAS.485.1293S}
{Schnitzeler}, D.~H.~F.~M., {Carretti}, E., {Wieringa}, M.~H., {et~al.} 2019,
  \mnras, 485, 1293

\bibitem[{{Shanahan} {et~al.}(2019){Shanahan}, {Lemmer}, {Stil}, {Beuther},
  {Wang}, {Soler}, {Anderson}, {Bigiel}, {Glover}, {Goldsmith}, {Klessen},
  {McClure-Griffiths}, {Reissl}, {Rugel}, \& {Smith}}]{2019ApJ...887L...7S}
{Shanahan}, R., {Lemmer}, S.~J., {Stil}, J.~M., {et~al.} 2019, \apjl, 887, L7

\bibitem[{{Short} {et~al.}(2007){Short}, {Higdon}, \&
  {Kronberg}}]{2007BayAn...2..665S}
{Short}, M.~B., {Higdon}, D.~M., \& {Kronberg}, Philipp, P. 2007, Bayesian
  Analysis, 2, 665

\bibitem[{{Simard-Normandin} {et~al.}(1981){Simard-Normandin}, {Kronberg}, \&
  {Button}}]{1981ApJS...45...97S}
{Simard-Normandin}, M., {Kronberg}, P.~P., \& {Button}, S. 1981, \apjs, 45, 97

\bibitem[{{Sobey} {et~al.}(2019){Sobey}, {Bilous}, {Grie{\ss}meier}, {Hessels},
  {Karastergiou}, {Keane}, {Kondratiev}, {Kramer}, {Michilli}, {Noutsos},
  {Pilia}, {Polzin}, {Stappers}, {Tan}, {van Leeuwen}, {Verbiest},
  {Weltevrede}, {Heald}, {Alves}, {Carretti}, {En{\ss}lin}, {Haverkorn},
  {Iacobelli}, {Reich}, \& {Van Eck}}]{2019MNRAS.484.3646S}
{Sobey}, C., {Bilous}, A.~V., {Grie{\ss}meier}, J.~M., {et~al.} 2019, \mnras,
  484, 3646

\bibitem[{{Stil} {et~al.}(2011){Stil}, {Taylor}, \&
  {Sunstrum}}]{2011ApJ...726....4S}
{Stil}, J.~M., {Taylor}, A.~R., \& {Sunstrum}, C. 2011, \apj, 726, 4

\bibitem[{{Subramanian}(2016)}]{2016RPPh...79g6901S}
{Subramanian}, K. 2016, Reports on Progress in Physics, 79, 076901

\bibitem[{{Tabara} \& {Inoue}(1980)}]{1980A&AS...39..379T}
{Tabara}, H. \& {Inoue}, M. 1980, \aaps, 39, 379

\bibitem[{{Taylor} {et~al.}(2009){Taylor}, {Stil}, \&
  {Sunstrum}}]{2009ApJ...702.1230T}
{Taylor}, A.~R., {Stil}, J.~M., \& {Sunstrum}, C. 2009, \apj, 702, 1230

\bibitem[{{The NIFTy5 team} {et~al.}(2019){The NIFTy5 team}, {Arras},
  {En{\ss}lin}, {Frank}, {Hutschenreuter}, {Knollmueller}, {Leike},
  {Newrzella}, {Platz}, {Reinecke}, \& {Stadler}}]{NIFTY}
{The NIFTy5 team}, {Arras}, Philipp snd~{Baltac}, M., {En{\ss}lin}, T.~A.,
  {et~al.} 2019, {In preparation}

\bibitem[{{Vall{\'e}e}(2017)}]{2017AstRv..13..113V}
{Vall{\'e}e}, J.~P. 2017, The Astronomical Review, 13, 113

\bibitem[{{van der Velden}(2020)}]{2020JOSS....5.2004V}
{van der Velden}, E. 2020, The Journal of Open Source Software, 5, 2004

\bibitem[{{Van Eck}(2018)}]{2018Galax...6..112V}
{Van Eck}, C. 2018, Galaxies, 6, 112

\bibitem[{{Van Eck} {et~al.}(submitted){Van Eck}, {Brown}, {Ordog}, \&
  et~al}]{van_Eck_cgps}
{Van Eck}, C., {Brown}, J., {Ordog}, A., \& et~al. submitted

\bibitem[{{Van Eck} {et~al.}(2011){Van Eck}, {Brown}, {Stil}, {Rae}, {Mao},
  {Gaensler}, {Shukurov}, {Taylor}, {Haverkorn}, {Kronberg}, \&
  {McClure-Griffiths}}]{2011ApJ...728...97V}
{Van Eck}, C.~L., {Brown}, J.~C., {Stil}, J.~M., {et~al.} 2011, \apj, 728, 97

\bibitem[{{Van Eck} {et~al.}(2018){Van Eck}, {Haverkorn}, {Alves}, {Beck},
  {Best}, {Carretti}, {Chy{\.z}y}, {Farnes}, {Ferri{\`e}re}, {Hardcastle},
  {Heald}, {Horellou}, {Iacobelli}, {Jeli{\'c}}, {Mulcahy}, {O'Sullivan},
  {Polderman}, {Reich}, {Riseley}, {R{\"o}ttgering}, {Schnitzeler}, {Shimwell},
  {Vacca}, {Vink}, \& {White}}]{2018A&A...613A..58V}
{Van Eck}, C.~L., {Haverkorn}, M., {Alves}, M.~I.~R., {et~al.} 2018, \aap, 613,
  A58

\bibitem[{{Van Eck et al.}(in prep.)}]{van_Eck_catalog}
{Van Eck et al.} in prep.

\bibitem[{{van Haarlem} {et~al.}(2013){van Haarlem}, {Wise}, {Gunst}, {Heald},
  {McKean}, {Hessels}, {de Bruyn}, {Nijboer}, {Swinbank}, {Fallows},
  {Brentjens}, {Nelles}, {Beck}, {Falcke}, {Fender}, {H{\"o}randel},
  {Koopmans}, {Mann}, {Miley}, {R{\"o}ttgering}, {Stappers}, {Wijers},
  {Zaroubi}, {van den Akker}, {Alexov}, {Anderson}, {Anderson}, {van Ardenne},
  {Arts}, {Asgekar}, {Avruch}, {Batejat}, {B{\"a}hren}, {Bell}, {Bell}, {van
  Bemmel}, {Bennema}, {Bentum}, {Bernardi}, {Best}, {B{\^\i}rzan}, {Bonafede},
  {Boonstra}, {Braun}, {Bregman}, {Breitling}, {van de Brink}, {Broderick},
  {Broekema}, {Brouw}, {Br{\"u}ggen}, {Butcher}, {van Cappellen}, {Ciardi},
  {Coenen}, {Conway}, {Coolen}, {Corstanje}, {Damstra}, {Davies}, {Deller},
  {Dettmar}, {van Diepen}, {Dijkstra}, {Donker}, {Doorduin}, {Dromer}, {Drost},
  {van Duin}, {Eisl{\"o}ffel}, {van Enst}, {Ferrari}, {Frieswijk}, {Gankema},
  {Garrett}, {de Gasperin}, {Gerbers}, {de Geus}, {Grie{\ss}meier}, {Grit},
  {Gruppen}, {Hamaker}, {Hassall}, {Hoeft}, {Holties}, {Horneffer}, {van der
  Horst}, {van Houwelingen}, {Huijgen}, {Iacobelli}, {Intema}, {Jackson},
  {Jelic}, {de Jong}, {Juette}, {Kant}, {Karastergiou}, {Koers}, {Kollen},
  {Kondratiev}, {Kooistra}, {Koopman}, {Koster}, {Kuniyoshi}, {Kramer},
  {Kuper}, {Lambropoulos}, {Law}, {van Leeuwen}, {Lemaitre}, {Loose}, {Maat},
  {Macario}, {Markoff}, {Masters}, {McFadden}, {McKay-Bukowski}, {Meijering},
  {Meulman}, {Mevius}, {Middelberg}, {Millenaar}, {Miller-Jones}, {Mohan},
  {Mol}, {Morawietz}, {Morganti}, {Mulcahy}, {Mulder}, {Munk}, {Nieuwenhuis},
  {van Nieuwpoort}, {Noordam}, {Norden}, {Noutsos}, {Offringa}, {Olofsson},
  {Omar}, {Orr{\'u}}, {Overeem}, {Paas}, {Pandey-Pommier}, {Pandey}, {Pizzo},
  {Polatidis}, {Rafferty}, {Rawlings}, {Reich}, {de Reijer}, {Reitsma},
  {Renting}, {Riemers}, {Rol}, {Romein}, {Roosjen}, {Ruiter}, {Scaife}, {van
  der Schaaf}, {Scheers}, {Schellart}, {Schoenmakers}, {Schoonderbeek},
  {Serylak}, {Shulevski}, {Sluman}, {Smirnov}, {Sobey}, {Spreeuw}, {Steinmetz},
  {Sterks}, {Stiepel}, {Stuurwold}, {Tagger}, {Tang}, {Tasse}, {Thomas},
  {Thoudam}, {Toribio}, {van der Tol}, {Usov}, {van Veelen}, {van der Veen},
  {ter Veen}, {Verbiest}, {Vermeulen}, {Vermaas}, {Vocks}, {Vogt}, {de Vos},
  {van der Wal}, {van Weeren}, {Weggemans}, {Weltevrede}, {White}, {Wijnholds},
  {Wilhelmsson}, {Wucknitz}, {Yatawatta}, {Zarka}, {Zensus}, \& {van
  Zwieten}}]{2013A&A...556A...2V}
{van Haarlem}, M.~P., {Wise}, M.~W., {Gunst}, A.~W., {et~al.} 2013, \aap, 556,
  A2

\bibitem[{{Vernstrom} {et~al.}(2018){Vernstrom}, {Gaensler}, {Vacca}, {Farnes},
  {Haverkorn}, \& {O'Sullivan}}]{2018MNRAS.475.1736V}
{Vernstrom}, T., {Gaensler}, B.~M., {Vacca}, V., {et~al.} 2018, \mnras, 475,
  1736

\bibitem[{{Wrobel}(1993)}]{1993AJ....106..444W}
{Wrobel}, J.~M. 1993, \aj, 106, 444

\bibitem[{{Wu} {et~al.}(2015){Wu}, {Kim}, \& {Ryu}}]{2015NewA...34...21W}
{Wu}, Q., {Kim}, J., \& {Ryu}, D. 2015, \na, 34, 21

\bibitem[{{Xu} \& {Han}(2014)}]{2014RAA....14..942X}
{Xu}, J. \& {Han}, J.-L. 2014, Research in Astronomy and Astrophysics, 14, 942

\bibitem[{{Xu} {et~al.}(2006){Xu}, {Kronberg}, {Habib}, \&
  {Dufton}}]{2006ApJ...637...19X}
{Xu}, Y., {Kronberg}, P.~P., {Habib}, S., \& {Dufton}, Q.~W. 2006, \apj, 637,
  19

\bibitem[{{Yao} {et~al.}(2017){Yao}, {Manchester}, \&
  {Wang}}]{2017ApJ...835...29Y}
{Yao}, J.~M., {Manchester}, R.~N., \& {Wang}, N. 2017, \apj, 835, 29

\end{thebibliography}

\end{document}